\documentclass[preprintnumbers, prd, showpacs, floatfix,  superscriptaddress,nofootinbib,notitlepage]{revtex4-1}

\usepackage{graphicx}
\usepackage{latexsym}
\usepackage{amssymb}
\usepackage{amsmath}

\usepackage{dsfont}
\usepackage{esint}

\usepackage{xcolor}
\usepackage[ocgcolorlinks,colorlinks=true,linkcolor=blue,citecolor=blue,urlcolor=blue,pdftex]{hyperref}

\newcommand{\hh}[1]{\overset{\scriptscriptstyle{(#1)}}{h}{}}
\newcommand{\PP}[1]{\overset{\scriptscriptstyle{(#1)}}{\Phi}{}}
\newcommand{\PPh}[1]{\overset{\scriptscriptstyle{(#1)}}{\hat{\Phi}}{}}

\newcommand{\hypgeo}[2]{\vphantom{F}_{#1}F_{#2}}

\begin{document}

\title{Post-Newtonian parameter $\gamma$ for multiscalar-tensor gravity with a general potential}
\author{Manuel Hohmann}
\email{manuel.hohmann@ut.ee}
\affiliation{Institute of Physics, University of Tartu, W.\ Ostwaldi 1, Tartu 50411, Estonia}
\author{Laur J\"arv}
\email{laur.jarv@ut.ee}
\affiliation{Institute of Physics, University of Tartu, W.\ Ostwaldi 1, Tartu 50411, Estonia}
\author{Piret Kuusk}
\email{piret.kuusk@ut.ee}
\affiliation{Institute of Physics, University of Tartu, W.\ Ostwaldi 1, Tartu 50411, Estonia}
\author{Erik Randla}
\email{erik.randla@ut.ee}
\affiliation{Institute of Physics, University of Tartu, W.\ Ostwaldi 1, Tartu 50411, Estonia}
\author{Ott Vilson}
\email{ovilson@ut.ee}
\affiliation{Institute of Physics, University of Tartu, W.\ Ostwaldi 1, Tartu 50411, Estonia}


\begin{abstract}
We compute the parametrized post-Newtonian parameter $\gamma$ in the case of a static point source for multiscalar-tensor gravity with completely general nonderivative couplings and potential in the Jordan frame. Similarly to the single massive field case $\gamma$ depends exponentially on the distance from the source and is determined by the length of a vector of non-minimal coupling in the space of scalar fields and its orientation relative to the mass eigenvectors. Using data from the Cassini tracking experiment, we estimate bounds on a general theory with two scalar fields.
Our formalism can be utilized for a wide range of models, which we illustrate by applying it to nonminimally coupled Higgs SU(2) doublet, general hybrid metric-Palatini gravity, linear ($\Box^{-1}$) and quadratic ($\Box^{-2}$) nonlocal gravity.
\end{abstract}

\maketitle

\nopagebreak

\section{Introduction}

Multiscalar-tensor gravity (MSTG) generalizes the scalar-tensor gravity (STG) of a scalar field $\Phi$ nonminimally coupled to curvature $R$, to the case of multiple scalar fields $\Phi^\alpha$ \cite{DamourEF,Berkin:1993bt}. Nonminimal couplings are typically generated by quantum corrections and arise in the effective models of higher dimensional theories. Diverse versions of MSTG appear in fundamental physics and cosmology in various constructions and under different disguises.

First, there are several phenomenological motivations to consider nonminimally coupled scalars. The Standard Model Higgs field is an SU(2) complex doublet, 
in the case it is endowed with a nonminimal coupling to curvature also the Goldstone modes may play a role in Higgs inflation \cite{Higgs inflation}
and the subsequent dark energy era \cite{Higgs dark energy}.
Otherwise a nonminimal Higgs may be paired with another nonminimal scalar (e.g.\ a dilaton) \cite{Higgs and scalar},
or the inflation and dark energy could be run by two nonminimally coupled  scalars \cite{biscalar models}.
More general MSTG inflation or dark energy models have $N$ fields with noncanonical kinetic terms and arbitrary potential \cite{DamourEF,MSTG inflation,Einstein frame metric,Kuusk:2014sna,meie2015} (also considered for stellar models \cite{Horbatsch:2015bua}), 
or are embedded into a supergravity setup \cite{Higgs sugra}.
The most general multiscalar-tensor gravitational action with second order field equations includes derivative couplings and is a generalization of Horndeski's class of theories, so far worked out for the two fields case \cite{biscalar Horndeski}.

Second, different proposed extensions and modifications of general relativity can be also cast into the form of MSTG by a change of variables. 
It is well-known that, if the gravitational lagrangian is nonlinear in curvature, $f(R)$, or more generally $f(\Phi^\alpha,R)$, the theory is dynamically equivalent to (M)STG with the potential depending on the form of the function $f$ \cite{f(R),f(Phi R),f(Phii R)}.
Likewise we get an MSTG when the original lagrangian is a more complicated function of multiple arguments of $R$, $\Box R$, $\nabla_\mu R \nabla^\mu R$, Gauss-Bonnet topological term, or Weyl tensor squared \cite{f(R G C etc}, as each such argument can contribute a scalar nonminimally coupled to $R$. 
(A function of arbitrary curvature invariants can be also turned into scalars, but the tensor part will not generally reduce to linear $R$ alone \cite{f(general R)}.)
If the metric and connection are treated as independent variables, defining curvature scalars $R$ and $\mathcal{R}$, the resulting general hybrid metric-Palatini $f(R,\mathcal{R})$ gravity is equivalent to MSTG with two scalars \cite{general hybrid metric-Palatini}.
A related construction called C-theory which continuously interpolates between metric and Palatini gravities, also possesses a biscalar-tensor representation \cite{C-theory}.
In case the lagrangian is a function of higher derivatives of the curvature, $f(R, \Box^i R)$, each such argument can be converted to a nonmimimal scalar in MSTG \cite{f(box^i R)}. 
Moreover, a lagrangian of nonlocal gravity, characterized by derivatives in the inverse powers, $f(R, \Box^{-i} R)$, can be made local by again introducing auxiliary scalar fields nonminimally coupled to $R$ \cite{nonlocal,sasaki nonlocal}.

The parametrized post-Newtonian (PPN) formalism is designed to describe slow motions in weak gravitational fields \cite{Will}, and can be utilized to confront the theory with high precision measurements in e.g.\ the Solar System.
The original STG computation by Nordtvedt \cite{Nordtvedt1970}, which assumed that the potential (mass) of the scalar field vanishes,
has been generalized to studies of higher order effects \cite{2PN} and for models with altered kinetic term or extra scalar-matter couplings \cite{PPN_nonstandard_STG_without_potential}. 
An important lesson learned in the STG case is that making the scalar field massive by the inclusion of the potential modifies the PPN parameters  \cite{Olmo_Perivolaropoulos,meie2013,Scharer:2014kya,meie2014}, so that the theory becomes viable in a much larger domain (cf.\ also Refs.\ \cite{massive_BD}, \cite{STG_neutron_stars}). 
This has been especially relevant for understanding the PPN behavior of $f(R)$ gravity \cite{ppn_fR}, equivalent to a subclass of STG. 
Of course, a similar effect is also present in the generalized STG or Horndeski theory \cite{Horndeski_ppn}.
Curiously enough, in teleparallel theories where the scalar field is nonminimally coupled to torsion, the PPN parameters coincide with the ones of general relativity \cite{teleparallel_ppn} unless a boundary term is introduced to the action \cite{teleparallel_boundary_ppn}. 

In the pioneering MSTG paper Damour and Esposito-Far\`ese derived the PPN parameters in the Einstein frame and assuming the potential vanishes \cite{DamourEF}. The effect of the potential was also not considered in the Jordan frame computation for the constant diagonal kinetic term \cite{Berkin:1993bt} and more recently for a generic kinetic term \cite{erik2014}. These results were generalized to arbitrary frame and scalar field parametrization by using the formalism of invariants \cite{meie2015}.
The PPN parameters for C-theory have been determined by relying on the correspondence with a subclass of MSTG in the massless (vanishing potential) limit \cite{C-theory PPN}, while the PPN parameter $\gamma$ for specific nonlocal gravity models has been found using the biscalar representation \cite{Koivisto:2008dh} as well as independently of MSTG \cite{Conroy:2014eja}.

The purpose of this paper is to calculate the PPN parameter $\gamma$ for a Jordan frame MSTG with generic kinetic terms and arbitrary potential (but without derivative couplings). In Sec.\ \ref{Sec_2} we recall the Jordan frame MSTG action in different parametrizations. In the process we define a covariant metric on the space of scalar fields and the vector of nonminimal coupling, these allow us to clarify the invariant notion of ghosts and the meaning of nonminimal coupling. Next in Sec.\ \ref{Sec_3} we carry out the PPN computation for a point mass source and find that the effective gravitational constant as well as the PPN parameter $\gamma$ in general depend on the distance from the source. Sec.\ \ref{Sec_4} uncovers the geometric picture underlying this result in terms of the eigenvectors of the mass matrix. Sec.\ \ref{Sec_5} draws rough experimental bounds on the two scalars case from the Cassini tracking experiment. Sec.\ \ref{Sec_6} illustrates how to apply our formalism for various interesting examples of MSTG: nonminimally coupled Higgs SU(2) doublet, general hybrid metric-Palatini gravity, linear ($\Box^{-1}$) nonlocal gravity, and quadratic ($\Box^{-2}$) nonlocal gravity. The last section \ref{Sec_7} provides a summary and outlook. 

Some more technical calculations are given in the appendices. Appendix \ref{App_A} discusses when the mass matrix can be diagonalized. Appendix \ref{App_B} deals with the boundary value problem and the determination of integration constants. Appendix \ref{App_C} addresses the cases when the mass matrix can not be brought into diagonal form and there are higher dimensional Jordan blocks.


\section{Jordan frame action functional for $N$ fields in different parametrizations \label{secAct}}
\label{Sec_2}

We start our discussion of multiscalar-tensor gravity with a brief review of its action functional and field equations in the Jordan frame. In section \ref{Subsec 2.1}, we consider a general parametrization of the scalars, while in section \ref{Subsec 2.2} we present the special case of a Brans-Dicke like parametrization.

\subsection{General parametrization}
\label{Subsec 2.1}

The general form of the multiscalar-tensor gravity action in the Jordan frame 
with $N$ scalar fields $\Phi^\alpha$ can be written as \cite{Berkin:1993bt,Kuusk:2014sna,erik2014}
\begin{eqnarray}
S =
\frac{1}{2\kappa^2}\int d^{4}{x}\sqrt{-g}\left(\mathcal{F} R - 
\mathcal{Z}_{\alpha \beta}g^{\mu\nu}\partial_\mu \Phi^\alpha \partial_\nu \Phi^\beta - 2\kappa^2 \mathcal{U}\right)+
S_m [g_{\mu\nu},\chi ] \,.
\label{eq:MSTG:Action_kuusk}
\end{eqnarray}
Here the indices $\alpha,\beta,\gamma, \ldots = 1,2,\ldots,N$ label the scalar fields, the indices $\mu,\nu,\ldots=0,1,2,3$ belong to the spacetime coordinates, while $i,j,\ldots=1,2,3$ are reserved for the spatial coordinates in the calculations later.
The function $\mathcal{F}=\mathcal{F}(\Phi^1,\Phi^2,\ldots,\Phi^N)>0$ describes the nonminimal coupling between the scalars and curvature, making the effective gravitational constant field dependent.
The functions $\mathcal{Z}_{\alpha \beta}=\mathcal{Z}_{\alpha \beta}(\Phi^1,\Phi^2,\ldots,\Phi^N)$ characterize the kinetic terms of the scalar fields, while $\mathcal{U}=\mathcal{U}(\Phi^1,\Phi^2,\ldots,\Phi^N)$ denotes the scalar potential. In the Jordan frame the action $S_m [g_{\mu\nu},\chi_m ]$ for the matter fields $\chi_m$ involves only the metric $g_{\mu \nu}$ and not the scalars $\Phi^\alpha$. Making a scalar fields dependent conformal rescaling of the metric will present the theory in a different frame where the matter action $S_m$ would contain the scalars as well.
Likewise we can also reparametrize the scalar fields, changing the form of the functions $\mathcal{F}$, $\mathcal{Z}_{\alpha \beta}$, $\mathcal{U}$ and possibly recasting the theory into a form more amenable for computations and physical interpretation.
We have adopted a system of units where the speed of light and Planck's constant are set to equal one, $c=h=1$. 
The constant $\frac{\kappa^2}{8\pi}$ can be interpreted as a bare Newtonian gravitational constant, while effectively the strength of gravity is modified by the function $\mathcal{F}$.

The variation of the action (\ref{eq:MSTG:Action_kuusk}) with respect to the metric gives a generalization of the Einstein's equation,
\begin{eqnarray}
  \mathcal{F} (R_{\mu\nu}- \frac{1}{2}g_{\mu\nu}R) + g_{\mu\nu} \Box \mathcal{F} - \nabla_{\mu}\nabla_{\nu}\mathcal{F}+ \frac{1}{2} g_{\mu\nu} \mathcal{Z}_{\alpha \beta} \nabla_{\rho}\Phi^\alpha \nabla^{\rho}\Phi^\beta - \mathcal{Z}_{\alpha \beta}  \nabla_{\mu}\Phi^\alpha \nabla_{\nu}\Phi^\beta + \kappa^2 g_{\mu\nu} \mathcal{U}  = \kappa^2 T_{\mu\nu}^{(\chi)} \,,
  \label{eq:MSTG:JFfeq1}
\end{eqnarray}
while the variation with respect of the scalar fields and eliminating the $R$ term by using the trace of Eq.\ (\ref{eq:MSTG:JFfeq1}) yields equations for the scalar fields,
\begin{eqnarray}
  \label{eq:MSTG:ScalarFEQ}
\Bigg( 2 \mathcal{F} \mathcal{Z}_{\alpha \beta} + 3 \frac{\partial \mathcal{F}}{\partial \Phi^\alpha} \frac{\partial \mathcal{F}}{\partial \Phi^\beta} \Bigg) \Box \Phi^\beta =  
- 3 \frac{\partial \mathcal{F}}{\partial \Phi^\alpha} \frac{\partial^2 \mathcal{F}}{\partial \Phi^\beta \partial \Phi^\delta} \partial_{\rho} \Phi^\beta \partial^{\rho} \Phi^\delta 
- \frac{\partial \mathcal{F}}{\partial \Phi^\alpha} \mathcal{Z}_{\beta \delta} \partial_{\rho} \Phi^\beta \partial^{\rho} \Phi^\delta +
\nonumber \\
\qquad \qquad \qquad
+ \mathcal{F} \frac{\partial \mathcal{Z}_{\beta \delta}}{\partial \Phi^\alpha} \partial_{\rho} \Phi^\beta \partial^{\rho} \Phi^\delta 
- 2 \mathcal{F} \frac{\partial \mathcal{Z}_{\alpha \beta}}{\partial \Phi^\delta} \partial_{\rho} \Phi^\beta \partial^{\rho} \Phi^\delta 
- 4 \frac{\partial \mathcal{F}}{\partial \Phi^\alpha} \kappa^2 \mathcal{U}
+ 2 \mathcal{F} \kappa^2 \frac{\partial \mathcal{U}}{\partial \Phi^\alpha}
+ \frac{\partial \mathcal{F}}{\partial \Phi^\alpha} \kappa^2 T^{(\chi)}\,.
\end{eqnarray}
The energy-momentum tensor of matter fields $\chi$ is defined by
\begin{equation}
T_{\mu\nu}^{(\chi)} = -\frac{2}{\sqrt{-g}} \frac{\delta S_m}{\delta g^{\mu \nu}} \,.
\end{equation}

It is well known that in the theories where the scalars are minimally coupled to gravity, the matrix $\mathcal{Z}_{\alpha \beta}$ can be interpreted as a metric for the space of scalar fields, it transforms as a second rank covariant tensor under the redefinitions of the scalar fields. However, when the scalars are nonminimally coupled to curvature, the matrix 
\begin{equation}
\mathcal{F}_{\alpha\beta} \equiv \frac{1}{4\mathcal{F}^2} \Bigg( 2 \mathcal{F} \mathcal{Z}_{\alpha\beta} + 3 \frac{\partial \mathcal{F}}{\partial \Phi^\alpha} \frac{\partial \mathcal{F}}{\partial \Phi^\beta} \Bigg) \,
\label{F_ab}
\end{equation}
is a natural candidate for the role of the metric of the space spanned by the scalar fields. It transforms covariantly not only under the redefinitions of the scalar fields, but also under the conformal rescalings of the spacetime metric \cite{meie2015}. For instance we can use it to define a scalar product of quantities which are vectors in the space of scalar fields. 

In a well behaving theory the eigenvalues of $\mathcal{F}_{\alpha\beta}$ are positive, while a negative eigenvalue signals the presence of a ghost among the scalars. The latter becomes quite apparent when we make a conformal transformation into the Einstein frame where the metric and scalar variables of gravitation are more clearly separated. In the Einstein frame $\mathcal{F}\equiv 1$ and $\mathcal{F}_{\alpha\beta}$ reduces to $\frac{1}{2}\mathcal{Z}_{\alpha\beta}$ which is the factor in front of the scalar kinetic terms. Therefore its negative eigenvalue implies a ``wrong'' sign kinetic term for one of the scalar degrees of freedom \cite{Einstein frame metric,sasaki nonlocal}. 

A zero eigenvalue of $\mathcal{F}_{\alpha\beta}$ tells that one of the scalar degrees of freedom is nondynamical (like $\omega_{BD}=-\frac{3}{2}$ for a single field Brans-Dicke case). In this case the equations of motion allow one of the fields to be expressed in terms of the other fields, and by inserting this relation into the action we can integrate the nondynamical field away (see e.g.\ Ref.\ \cite{C-theory PPN}). Note that zero eigenvalue implies that the metric $\mathcal{F}_{\alpha\beta}$ is not invertible and the computation scheme developed in the current paper does not go through.

We assume the matrix $\mathcal{F}_{\alpha\beta}$ has an inverse $\mathcal{F}^{\beta\gamma}$, i.e.\ $\mathcal{F}_{\alpha\beta} \mathcal{F}^{\beta\gamma} = \delta_\alpha^\gamma$ and $\det(\mathcal{F}_{\alpha\beta}) \neq 0$. 
Then we may multiply the scalar field equations (\ref{eq:MSTG:ScalarFEQ}) with $\mathcal{F}^{\gamma\alpha}$ from the left, to establish
\begin{equation}
\label{eq:kastfi}
\Box\Phi^\gamma=\mathcal{E}^{\gamma} - \mathcal{K}^\gamma T^{(\chi)}  \,,\end{equation}
where the kinetic and potential terms are collected into
\begin{eqnarray}
\mathcal{E}^\gamma&=& \mathcal{F}^{\gamma\alpha} \Big( - \frac{3}{4\mathcal{F}^2} \frac{\partial \mathcal{F}}{\partial \Phi^\alpha} \frac{\partial^2 \mathcal{F}}{\partial \Phi^\beta \partial \Phi^\delta} \partial_{\rho} \Phi^\beta \partial^{\rho} \Phi^\delta 
- \frac{1}{4\mathcal{F}^2} \frac{\partial \mathcal{F}}{\partial \Phi^\alpha} \mathcal{Z}_{\beta\delta} \partial_{\rho} \Phi^\beta \partial^{\rho} \Phi^\delta 
\nonumber \\
 && \qquad \quad
+ \frac{1}{4\mathcal{F}} \frac{\partial \mathcal{Z}_{\beta\delta}}{\partial \Phi^\alpha} \partial_{\rho} \Phi^\beta \partial^{\rho} \Phi^\delta
- \frac{1}{2\mathcal{F}} \frac{\partial \mathcal{Z}_{\alpha\beta}}{\partial \Phi^\delta} \partial_{\rho} \Phi^\beta \partial^{\rho} \Phi^\delta 
-{ \frac{1}{\mathcal{F}^2} \frac{\partial\mathcal{F}}{\partial \Phi^\alpha} \kappa^2 \mathcal{U}
+ \frac{\kappa^2}{2\mathcal{F}}  \frac{\partial\mathcal{U}}{\partial \Phi^\alpha}} \Big)\,,
\end{eqnarray}
and the influence of matter energy-momentum is mediated by
\begin{equation}
\mathcal{K}_\alpha = -\kappa^2 \frac{1}{4\mathcal{F}^2} \frac{\partial \mathcal{F}}{\partial \Phi^\alpha}  \,, \qquad \quad
\mathcal{K}^\gamma = -\kappa^2 \frac{1}{4\mathcal{F}^2} \mathcal{F}^{\gamma \alpha} \frac{\partial \mathcal{F}}{\partial \Phi^\alpha}  \,.
\label{k def}
\end{equation}

We may call the object with components $\mathcal{K}^\gamma$ a vector of nonminimal coupling since it is constructed from the derivatives of the nonminimal coupling function $\mathcal{F}$ and it transforms as a vector under the scalar field redefinitions. 
In MSTG the gravitational interaction is mediated by the metric and the nonminimal scalars: 
the dynamical equation for the metric  (\ref{eq:MSTG:JFfeq1}) is sourced by the energy-momentum tensor of the matter fields, $T^{(\chi)}_{\mu\nu}$, while the dynamics of the scalars (\ref{eq:kastfi}) is sourced by the trace of the matter energy-momentum, $T^{(\chi)}$.
If $\mathcal{K}_\alpha$ has a zero component then in this particular parametrization of the scalar fields the respective $\Phi^\alpha$ field is not directly coupled to the curvature. Analogously, if $\mathcal{K}^\gamma$ has a zero component, the field $\Phi^\gamma$ is not sourced by the matter. However, through the interactions between the scalars as encoded in the potential and kinetic terms of the action all scalars will still be indirectly coupled to curvature and matter in general. Only if all components of the vector of nonminimal coupling are identically zero, can we deem that the scalars as a collection are minimally coupled. For theories with positive definite metric $\mathcal{F}_{\alpha \beta}$ on the space of the scalar fields, if the length of the nonminimal coupling vector,
\begin{equation}
|\boldsymbol{\mathcal{K}}|^2 = \mathcal{F}_{\alpha\beta} \mathcal{K}^\alpha \mathcal{K}^\beta \,,
\label{k squared}
\end{equation}
vanishes, then the scalars are minimally coupled. The latter statement is frame and parametrization independent since the combination (\ref{k squared}) remains invariant under the redefinitions of the scalar fields as well as under the rescalings of the spacetime metric  \cite{meie2015}.


\subsection{Brans-Dicke like parametrization}
\label{BD_like_parametrization_sec}
\label{Subsec 2.2}

For a more straightforward physical interpretation of the theory it is convenient to redefine the scalar fields $\Phi^\alpha = \Phi^\alpha(\phi^1,\phi^2,\ldots,\phi^{N-1}, \Psi)$ by
setting \cite{Kuusk:2014sna,erik2014}
\begin{eqnarray}
\Psi = \mathcal{F}(\Phi^1,\Phi^2,\ldots,\Phi^N) \,. 
\label{Psi redefinition}
\end{eqnarray}
This reshuffles the scalars so that the covector of nonminimal coupling $\mathcal{K}_\alpha$ has only one nonzero component, i.e.\ it is aligned along the $N$th axis ($\Psi$ direction) in the space of the scalar fields. Taking into account that 
\begin{eqnarray}
  \label{eq:z} \hspace{-0.6cm}\mathcal{Z}_{\alpha \beta}\partial_\rho \Phi^\alpha \partial^\rho \Phi^\beta= \mathcal{Z}_{\alpha \beta}\left(\frac{\partial\Phi^\alpha}{\partial\phi^a} \frac{\partial\Phi^\beta}{\partial\phi^b}\partial_{\rho}\phi^a\partial^{\rho}\phi^b + 
2 \frac{\partial\Phi^\alpha}{\partial\phi^a} \frac{\partial\Phi^\beta}{\partial\Psi}\partial_{\rho}\phi^a\partial^{\rho}\Psi+ 
\frac{\partial\Phi^\alpha}{\partial\Psi} \frac{\partial\Phi^\beta}{\partial\Psi}\partial_{\rho}\Psi\partial^{\rho}\Psi\right)\,,
\end{eqnarray}
let us denote in this parametrization
\begin{eqnarray}
  \label{eq:redef}
Z_{ab}&=&\mathcal{Z}_{\alpha \beta}\frac{\partial\Phi^\alpha}{\partial\phi^a} \frac{\partial\Phi^\beta}{\partial\phi^b} \,, \nonumber\\
Z_{aN}&=&\mathcal{Z}_{\alpha \beta}\frac{\partial\Phi^\alpha}{\partial\phi^a} \frac{\partial\Phi^\beta}{\partial\Psi} \,, \\
Z_{NN}&=&\mathcal{Z}_{\alpha \beta}\frac{\partial\Phi^\alpha}{\partial\Psi} \frac{\partial\Phi^\beta}{\partial\Psi} \,, \nonumber\\
U(\phi^1,\phi^2,\ldots,\phi^{N-1},\Psi)&=&\mathcal{U}(\Phi^1,\Phi^2,\ldots,\Phi^N)\,, \nonumber
\end{eqnarray}
where $a, b, \ldots = 1,2,\ldots, N - 1$ label the scalar fields $\phi$. 
In general there are all together $N$ scalar redefinitions (relations between the old and new set of scalar fields) at our disposal, one of these has been already employed as (\ref{Psi redefinition}). Let us use the remaining  $N-1$ conditions to impose
\begin{eqnarray}
  \label{eq:assume} 
  Z_{aN} (\phi^1,\phi^2,\ldots,\phi^{N-1},\Psi)=0\,.
\end{eqnarray}
We may also denote
\begin{eqnarray}
Z_{NN}= \frac{\omega(\phi^1,\phi^2,\ldots,\phi^{N-1},\Psi)}{\Psi}  \,. 
\end{eqnarray}
After such redefinitions all but one of the vector components $\mathcal{K}^\alpha$ vanish, and the action  (\ref{eq:MSTG:Action_kuusk}) reads
\begin{eqnarray}
  \label{eq:MSTG:ActionStar_kuusk}
  S = 
\frac{1}{2\kappa^2}\int d^{4}{x}\sqrt{-g}\left(\Psi R - 
Z_{ab}\partial_\rho \phi^a \partial^\rho \phi^b- 
\frac{\omega}{\Psi}\partial_{\rho}\Psi\partial^{\rho}\Psi - 2\kappa^2 U\right)+S_m [g_{\mu\nu},\chi_m ] \,,
\end{eqnarray}
looking akin to the Brans-Dicke theory. 
However note that the functions depend on all scalar fields: $\omega=\omega(\phi^1,\phi^2,\ldots,\Psi)$,
${Z}_{ab}={Z}_{ab}(\phi^1,\phi^2,\ldots,\Psi)$,
${U}=U(\phi^1,\phi^2,\ldots,\Psi)$. The scalar field $\Psi$ has been singled out as a variable part of the gravitational constant.
The field equations corresponding to the action (\ref{eq:MSTG:ActionStar_kuusk}) are
\begin{eqnarray}
  \Psi (R_{\mu\nu}- \frac{1}{2}g_{\mu\nu} R) + g_{\mu\nu} \Box \Psi - \nabla_{\mu}\nabla_{\nu}\Psi 
 + \frac{1}{2} g_{\mu\nu} \mathcal{Z}_{ab} \nabla_{\rho}\phi^a \nabla^{\rho}\phi^b 
+ \frac{1}{2} g_{\mu\nu} \frac{\omega}{\Psi} \nabla_{\rho}\Psi \nabla^{\rho}\Psi   
 &&
 \nonumber \\
 - \mathcal{Z}_{ab} \nabla_{\mu}\phi^a \nabla_{\nu}\phi^b
 - \frac{\omega}{\Psi} \nabla_{\mu}\Psi \nabla_{\nu}\Psi
 + \kappa^2 g_{\mu\nu} U  &=& \kappa^2 T_{\mu\nu}^{(\chi)} \,,
  \label{eq:feq1}
\end{eqnarray}
and
\begin{eqnarray}
  \label{eq:feq2}
  (2\omega +3)\Box \Psi &=& \left(\Psi\frac{\partial Z_{ab}}{\partial\Psi}-
  Z_{ab}\right)\partial_\rho\phi^a\partial^\rho \phi^b-\frac{\partial \omega}
  {\partial\Psi}\partial_\rho\Psi\partial^\rho \Psi - 2\frac{\partial \omega}
  {\partial\phi^a}\partial_\rho\phi^a\partial^\rho \Psi \nonumber \\ &&-2\kappa^2\left( 2U - \Psi \frac{\partial U}{\partial \Psi} - \frac{1}{2}T^{(\chi)} \right) \,, \\  
Z_{ac}\Box \phi^a &=&\left(\frac{1}{2}\frac{\partial Z_{ab}}{\partial \phi^c}-\frac{\partial Z_{ac}}{\partial 
\phi^b}\right)\partial_\rho\phi^a\partial^\rho \phi^b + \frac{1}{2\Psi}\frac{\partial \omega}{\partial 
\phi^c}\partial_\rho\Psi\partial^\rho \Psi 
-  \frac{\partial Z_{ac}}{\partial \Psi}\partial_\rho\phi^a\partial^\rho 
\Psi + \kappa^2 \frac{\partial U}{\partial \phi^c}\,. \label{eq:feq3}
\end{eqnarray}
It is clear that only the $\Psi$ field is directly sourced by the matter now.



\section{Computing the PPN parameter $\gamma$ for $N$ fields}
\label{Sec_3}

We now discuss the parametrized post-Newtonian (PPN) limit of the class of multiscalar-tensor theories introduced in the previous section. We start with a brief review of the PPN formalism for a single point mass source in the context of MSTG in section \ref{Sec_3.1}. Then we calculate the PPN limit starting with the zeroth velocity order in section \ref{Sec_3.2}, followed by the second velocity orders of the scalar field in section \ref{Sec_3.3}, the temporal metric component in section \ref{Sec_3.4} and its spatial components in section \ref{Sec_3.5}. Section \ref{Sec_3.6} confirms that in the case of a single scalar field we re-obtain the previously derived result.

\subsection{Post-Newtonian approximation}
\label{PPNapprox}
\label{Sec_3.1}

The PPN formalism has been developed to extract standardized information -- the PPN parameters -- characteristic of the slow motion weak field regime of metric gravity theories. We follow the customary PPN calculation procedure \cite{Will}. 

Matter is modeled by a perfect fluid whose stress-energy tensor is given by
\begin{eqnarray}
 \label{eqn:tmunu}
T^{\mu\nu} = (\rho + \rho\Pi + p)u^{\mu}u^{\nu} + pg^{\mu\nu}\,.
\end{eqnarray}
Here \(\rho\) is the rest energy density,  \(\Pi\) is the specific internal energy, \(p\) is the pressure and \(u^{\mu}\) is the four-velocity of matter. 
The gravitational field is assumed to be quasi-static, so that changes are only induced by the motion of the source matter. 
The orders of magnitude are ascribed to all quantities relative to the velocity \(v^{i} = u^{i}/u^0\) of the source matter, which is taken to be a first order small quantity:  $\rho \propto \Pi \propto p/\rho \propto v^2 \propto \mathcal{O}(2)$. Time derivatives of the metric components and the scalars 
are weighted with an additional velocity order $ \mathcal{O}(1)$. Later in the calculation we specify the matter source to be a point mass $M_0$ residing at the origin of spatial coordinates, $\rho=M_0 \delta(r)$.

The spacetime metric is taken to be a perturbed Minkowski metric $g_{\mu\nu}=\eta_{\mu\nu}+h_{\mu\nu}$. Only the metric components of order $\mathcal{O}(2)$, written as 
\begin{eqnarray}
\label{eq:h00}
\hh{2}_{00}&=&2 \, G_{\mathrm{eff}} \, \mathrm{U}_{\mathrm{N}}(r) \,, \\
\hh{2}_{ij}&=&2 \, G_{\mathrm{eff}} \, \gamma \, \mathrm{U}_{\mathrm{N}}(r) \, \delta_{ij}\,, \label{eq:hij}
\end{eqnarray}
are relevant for the calculation of the PPN parameter $\gamma$. Here $G_{\mathrm{eff}}$ is the effective gravitational constant and $\mathrm{U}_N(r)=\frac{M_0}{r}$ is the Newtonian gravitational potential which depends on the distance from the source point mass.


The temporal and spatial components of the Ricci tensor can be computed from their definition using the perturbed metric. Up to order $\mathcal{O}(2)$ they are
\begin{eqnarray}
\label{eq:Riccidef_00}
R_{00}&=& - \frac{1}{2} \nabla^2 \hh{2}_{00} \,, \\\label{eq:Riccidef}
R_{ij}&=& - \frac{1}{2} \nabla^2 \hh{2}_{ij} + \frac{1}{2} \left(\hh{2}_{00,ij}-\hh{2}_{kk,ij}+2\hh{2}_{ki,kj}\right)\,.
\end{eqnarray}
On the other hand, the equation of motion for the metric (\ref{eq:MSTG:JFfeq1}) can be trace-reversed to express the Ricci tensor components as
\begin{eqnarray}
R_{\mu\nu}=\frac{1}{\mathcal{F}}\left[ \kappa^2\left(T_{\mu\nu}-\frac{1}{2}g_{\mu\nu}T\right) + \kappa^2 g_{\mu\nu} \mathcal{U} + g_{\mu\nu}\Box\mathcal{F} -\frac{1}{2}g_{\mu\nu}\nabla^\rho \nabla_\rho \mathcal{F} + \nabla_\mu \nabla_\nu \mathcal{F} + \mathcal{Z}_{\alpha\beta}\nabla_\mu\Phi^\alpha \nabla_\nu\Phi^\beta \right] .\label{eq:RicciMSTG}
\end{eqnarray}
Equating the respective components of Eqs.\ (\ref{eq:Riccidef_00}) and (\ref{eq:Riccidef}) with those of (\ref{eq:RicciMSTG}), and solving for the metric components (\ref{eq:h00}) and (\ref{eq:hij}) is at the heart of the PPN $\gamma$ calculation.

All scalar fields are considered to be perturbed around their constant cosmological background values,
\begin{eqnarray}
{\Phi}^\alpha(x^\mu)=\PP{0}^\alpha+\PP{2}^\alpha(x^\mu)\,.\label{eq:scfpert}
\end{eqnarray}
In the asymptotic limit $\PP{2}^\alpha|_{r\rightarrow \infty} = 0$. The functions of the scalar fields are expanded in a Taylor series with the coefficients assumed to be of order $\mathcal{O}(0)$.

\subsection{$0^{\textrm{th}}$ order approximation}
\label{0th_order_section}
\label{Sec_3.2}

In the lowest order of magnitude $\mathcal{O}(0)$, the background is Minkowski space empty of matter and the scalars are constant. The field equation for the metric (\ref{eq:RicciMSTG}) reduces to
\begin{equation}
\kappa^2\eta_{\mu\nu}\left.\mathcal{U}\right|_0=0\,.
\end{equation}
Thus the asymptotic background value of the potential must be negligible (which is consistent with the situation in e.g.\ the Solar System), 
\begin{equation}
{\mathcal{U}}|_0  = 0 \,.
\label{U 0}
\end{equation}
Similarly, the scalar fields' equations (\ref{eq:MSTG:ScalarFEQ}) or (\ref{eq:kastfi}) in the lowest order of magnitude become 
\begin{equation}
 2 \left.\frac{\partial \mathcal{F}}{\partial \Phi^\gamma}\right|_0  \left.\mathcal{U}\right|_0 = \left.\mathcal{F}\right|_0  \left.\frac{\partial \mathcal{U}}{\partial \Phi^\gamma}\right|_0\,,
\end{equation}
leading to
\begin{equation}
\left.\frac{\partial \mathcal{U}}{\partial \Phi^\gamma}\right|_0=0\,,
\label{U deriv 0}
\end{equation}
if we assume $\mathcal{F}_0 \equiv \left.\mathcal{F}\right|_0\neq0$ (the gravitational constant at the cosmological background is not infinite).


\subsection{Scalar fields at $2^\textrm{nd}$ order}
\label{Sec_3.3}

The next step is to expand the scalar equations (\ref{eq:kastfi}) in Taylor series up to the order of magnitude $\mathcal{O}(2)$. Time derivatives and squares of spatial derivatives drop out, while the conditions (\ref{U 0}), (\ref{U deriv 0}) simplify the result to
\begin{equation}
\nabla^2 \PP{2}^\gamma= \mathcal{M}^\gamma_{\;\;\;\alpha}\PP{2}^\alpha + k^\gamma\rho \,,
\label{varphieq}
\end{equation}
where the vector of nonminimal coupling (\ref{k def}) is taken at the asymptotic background,
\begin{equation}
k^\gamma = \mathcal{K}^\gamma|_0 \,,
\end{equation}
and the components of the ``mass matrix'' are 
\begin{equation}
\label{mass_matrix}
\mathcal{M}^\gamma_{\;\;\;\alpha}=  \left[
\frac{\kappa^2}{2\mathcal{F}}   \, \mathcal{F}^{\gamma\beta} \frac{\partial^2 \mathcal{U}}{\partial \Phi^\beta \partial \Phi^\alpha } \right]_0 \,.
\end{equation}

It is easier to integrate Eq.\ (\ref{varphieq}) when the mass matrix is turned into its Jordan normal form, $J^{(\beta)}_{\;\;\;\;(\delta)}=(P^{-1})^{(\beta)}_{\;\;\;\;\gamma} \mathcal{M}^\gamma_{\;\;\;\alpha} P^\alpha_{\;\;\;(\delta)}$. 
Here the similarity matrix $\boldsymbol{P}$ is constructed from the components of the eigenvectors or generalized eigenvectors of the mass matrix. 
In general, the eigenvalues of the mass matrix can be complex and the Jordan normal form consists of Jordan blocks of the size that depends on the difference between the algebraic and geometric multiplicity of the respective eigenvalue.
However, as discussed in the next Sec.\  \ref{section_geometric_interpretation}, in the physically most well behaved case when the background metric of the space of scalar fields, $\mathcal{F}^{\gamma\beta}|_0$, is positive definite (hence there are no ghosts), and the Hessian of the potential, $\frac{\partial^2 \mathcal{U}}{\partial \Phi^\beta \partial \Phi^\alpha }|_0$, is positive semidefinite (hence the field configuration is not unstable), 
the mass matrix $\boldsymbol{\mathcal{M}}$ will have nonnegative  eigenvalues $m_{[\delta]}^2$ with equal algebraic and geometric multiplicity.
 In this case $\boldsymbol{J}$ is a diagonal matrix of the eigenvalues $m_{[\delta]}^2$, while the similarity matrix element $P^\alpha_{\;\;\;(\delta)}$ is the $\alpha$-th component of the $\delta$-th eigenvector, $v^\alpha_{\;\;(\delta)}$. 
The question of when the mass matrix can be diagonalized is discussed further in Appendix \ref{mass_matrix_eigenvalues}, and the generic case with nontrivial Jordan blocks is treated in Appendix \ref{higer_dim_Jordan_blocks}.

Here let us proceed with the assumption that $\boldsymbol{J}$ is diagonal with nonnegative entries.
Since the matrices $\boldsymbol{\mathcal{M}}$,  $\boldsymbol{P}$, and $\boldsymbol{J}$ are constant,
we can write Eq.\ (\ref{varphieq}) as
\begin{eqnarray}
\nabla^2 (P^{-1})^{(\beta)}_{\;\;\;\;\gamma}\PP{2}^\gamma=J^{(\beta)}_{\;\;\;\;(\gamma)}(P^{-1})^{(\gamma)}_{\;\;\;\;\alpha}\PP{2}^\alpha  +(P^{-1})^{(\beta)}_{\;\;\;\;\gamma} \, k^\gamma\rho \,.
\end{eqnarray}
In essence we have made a transformation in the scalar fields space from a generic basis into a basis given by the mass matrix eigenvectors, indices in brackets enumerate components in the mass eigenbasis.
In the new diagonal basis
\begin{eqnarray}
\PPh{2}^{(\beta)}&=&(P^{-1})^{(\beta)}_{\;\;\;\gamma}\PP{2}^\gamma\,,\\
\hat{k}^{(\beta)} &=& (P^{-1})^{(\beta)}_{\;\;\;\gamma} \, k^\gamma \,,
\end{eqnarray}
the equation for the scalar fields assumes the generic form of a screened Poisson equation
\begin{equation}
\label{Poisson screened nonnegative distinct eigenvalue}
\nabla^2\PPh{2}^{(\beta)}- m_{[\beta]}^2 \PPh{2}^{(\beta)} =\hat{k}^{(\beta)} \rho \,,
\end{equation}
since $J^{(\beta)}_{\;\;\;(\gamma)}=m^2_{[\beta]} \delta^{(\beta)}_{(\gamma)}$. 
Here the square brackets denote that the lower index $_{[\beta]}$ comes in pair with upper $^{(\beta)}$ and there is no summation.
Eq.\ (\ref{Poisson screened nonnegative distinct eigenvalue}) is solved by
\begin{eqnarray}
\PPh{2}^{(\beta)} = -\int d^3{r^\prime}\frac{e^{-m_{[\beta]}|\mathbf{r}-\mathbf{{r^\prime}}|}}{4\pi|\mathbf{r}-\mathbf{{r^\prime}}|}\hat{k}^{(\beta)} \rho\,.
\end{eqnarray} 
Substituting a point mass $M_0$ at the origin for the matter density distribution $\rho$ allows us to express the solution as 
\begin{eqnarray}
\PPh{2}^{(\beta)} = -\frac{M_0}{4\pi r} e^{-m_{[\beta]} r} \, \hat{k}^{(\beta)}\,.
\end{eqnarray} 
The integration constants have been fixed by demanding that $\PPh{2}^{(\beta)}$ vanish at spatial infinity and the source  matches the surrounding field (fulfills Gauss' theorem, see Appendix \ref{App_B}). 
To obtain the solutions for the original scalar fields we have to transform back, 
\begin{eqnarray}
\label{Phi2_solution}
\PP{2}^\alpha=P^\alpha_{\;\;\;(\beta)}\PPh{2}^{(\beta)}=-\frac{M_0}{4\pi r}
P^\alpha_{\;\;\;(\beta)}
E^{(\beta)}_{\;\;\;(\delta)} (P^{-1})^{(\delta)}_{\;\;\;\gamma}k^\gamma\,,
\end{eqnarray}
where the radius dependence is encoded in the matrix
\begin{equation}
\label{matrix_E_def}
E^{(\beta)}_{\;\;\;(\delta)} = \left(e^{-\sqrt{J}r}\right)^{(\beta)}_{\;\;\;(\delta)} = e^{-m_{[\delta]}r} \delta^{(\beta)}_{\;(\delta)} \,.
\end{equation}
As laid out systematically in Appendices \ref{App_B} and \ref{higer_dim_Jordan_blocks} the basic form of the solution (\ref{Phi2_solution}) also holds in the general case of arbitrary eigenvalues of the mass matrix, while the matrix (\ref{matrix_E_def}) is not necessarily diagonal, but gets adjusted to encompass complex eigenvalues and higher dimensional Jordan blocks (\ref{final_formula}).

\label{subsec:scalo2}


\subsection{Metric perturbation $h_{00}$ and $G_{\mathrm{eff}}$}
\label{Sec_3.4}

The equation for the temporal components of the spacetime metric is obtained from Eq.\ (\ref{eq:RicciMSTG}) by Taylor expanding up to the second order and taking into account Eq.\ (\ref{eq:Riccidef_00}) as well as the other relevant conditions from Secs.\ \ref{PPNapprox} and \ref{0th_order_section}. The result is
\begin{eqnarray}
\nabla^2\left(\hh{2}_{00}-\left.\frac{1}{\mathcal{F}_0}\frac{\partial\mathcal{F}}{\partial\Phi^\alpha}\right|_0\PP{2}^\alpha\right)=-\frac{\kappa^2}{\mathcal{F}_0}\rho\,.
\end{eqnarray}
We should also substitute in the $\PP{2}$ solution (\ref{Phi2_solution}) and keep in mind that the matter density $\rho$ is given by a point mass $M_0$ at the origin.
The solution for the components of the metric is
\begin{eqnarray}
\hh{2}_{00}=\frac{M_0}{4\pi r \mathcal{F}_0 }\left(\kappa^2-\left.\frac{\partial\mathcal{F}}{\partial\Phi^\alpha}\right|_0 P^\alpha_{\;\;\;(\beta)} 
E^{(\beta)}_{\;\;\;(\delta)} (P^{-1})^{(\delta)}_{\;\;\;\gamma}k^\gamma\right)\,.
\end{eqnarray}
By comparing with Eq.\ (\ref{eq:h00}) we can read off the effective gravitational constant as
\begin{eqnarray}
\label{G_eff}
G_{\mathrm{eff}}=\frac{\kappa^2}{8\pi \mathcal{F}_0}\left(1 - \Gamma(r) \right)\,.
\end{eqnarray}
where the deviation from the background value, 
\begin{subequations}
\label{Gamma_def}
\begin{eqnarray}
\label{Gamma_def_F}
\Gamma(r)&=& -\frac{1}{4 \mathcal{F}_0^2} \left[ \frac{\partial\mathcal{F}}{\partial\Phi^\alpha} P^\alpha_{\;\;\;(\beta)} 
E^{(\beta)}_{\;\;\;(\delta)} (P^{-1})^{(\delta)}_{\;\;\;\gamma} \mathcal{F}^{\gamma \varepsilon} \frac{\partial \mathcal{F}}{\partial \Phi^{\varepsilon}} \right]_0 \\
&=& -\frac{4 \mathcal{F}_0^2}{\kappa^4} \left[ k_\alpha P^\alpha_{\;\;\;(\beta)} 
E^{(\beta)}_{\;\;\;(\delta)} (P^{-1})^{(\delta)}_{\;\;\;\gamma} k^{\gamma} \right] \,,
\label{Gamma_def_k}
\end{eqnarray}
\end{subequations}
is expressed by using the definition of the vector of nonminimal coupling, Eq.\ (\ref{k def}).


\subsection{Metric perturbation $h_{ij}$ and the PPN parameter $\gamma$}
\label{Sec_3.5}

The equation for the spatial components of the spacetime metric is likewise obtained from Eq.\ (\ref{eq:RicciMSTG}) by Taylor expanding up to the second order and taking into account Eq.\ (\ref{eq:Riccidef}) along with the other relevant conditions from Secs.\ \ref{PPNapprox} and \ref{0th_order_section}. Imposing also the gauge conditions
\begin{eqnarray}
\frac{1}{2}\hh{2}_{00,ij}-\frac{1}{2}\hh{2}_{kk,ij}+\hh{2}_{ki,kj}=\frac{1}{\mathcal{F}_0}\left.\frac{\partial\mathcal{F}}{\partial\Phi^\alpha}\right|_0\PP{2}^\alpha_{,ij}\,
\end{eqnarray}
simplifies the resulting equation to  
\begin{eqnarray}\label{eq:meqTaylor}
\nabla^2\left(\hh{2}_{ij}+\left.\frac{1}{\mathcal{F}_0}\delta_{ij}\frac{\partial\mathcal{F}}{\partial\Phi^\alpha}\right|_0\PP{2}^\alpha\right)=-\frac{\kappa^2}{\mathcal{F}_0}\delta_{ij}\rho\,.
\end{eqnarray}
Substituting in the $\PP{2}$ solution (\ref{Phi2_solution}) and a point mass source leads to the solution
\begin{eqnarray}
\hh{2}_{ij}=\frac{\kappa^2 M_0 \delta_{ij}}{4\pi r \mathcal{F}_0}\left(1 + \Gamma(r) \right)\,.
\end{eqnarray}
Comparison with the definitions (\ref{eq:hij}) and (\ref{G_eff}) allows to deduce the PPN parameter $\gamma$ to be 
\begin{eqnarray}\label{gengamma}
\gamma=
\frac{1 + \Gamma(r)}
{1- \Gamma(r)}\,.
\end{eqnarray}

Deviation from the general relativity value $\gamma=1$ is set by a rather complicated term (\ref{Gamma_def}) that involves the scalar field masses as well as nonminimal coupling. If the masses are zero the dependence on the distance drops out, but the deviation term $\Gamma$ will only disappear in the limit when the (length of the) vector of nonminimal coupling (\ref{k squared}) also vanishes. Some insights concerning the geometry behind this formula are discussed further in Sec.\ \ref{geometric_interpretation_sec}.

\subsection{Single field case}
\label{Sec_3.6}

In the special case of a single scalar-tensor gravity 
$\mathcal{F}$, $\mathcal{Z}$, and $\mathcal{U}$ depend only on one field. The metric of the space of the scalar fields reduces to a single component $\mathcal{F}_{11} = \frac{1}{4\mathcal{F}^2} \left( 2\mathcal{F}\mathcal{Z}+3 \left( \frac{\partial \mathcal{F}}{\partial\Phi} \right)^2  \right)$
while the mass matrix is automatically diagonal,
\begin{eqnarray}
\mathcal{M}^1_{\;\;\;1}=
\left[\frac{2\kappa^2 \mathcal{F}}{2\mathcal{F}\mathcal{Z}+3 \left( \frac{\partial \mathcal{F}}{\partial\Phi} \right)^2 }\, \frac{\partial^2\mathcal{U}}{\partial\Phi^2} \right]_0
\equiv J^{(1)}_{\;\;\;(1)} \,.
\end{eqnarray}
The deviation term (\ref{Gamma_def}) is thus
\begin{equation}
\Gamma(r) = \left[ -\frac{1}{2\mathcal{F}\mathcal{Z}+3 \left( \frac{\partial \mathcal{F}}{\partial\Phi} \right)^2 } 
\left( \frac{\partial \mathcal{F}}{\partial\Phi} \right)^2 \right]_0
e^{-\sqrt{J}r}  \,.
\end{equation}

In the Brans-Dicke like parametrization where $\mathcal{F}=\Psi$ and $\mathcal{Z}=\frac{\omega}{\Psi}$, the metric simplifies to $\mathcal{F}_{\alpha \gamma} = \frac{1}{4\Psi^2} (2\omega+3)$,
and the mass matrix is given by
\begin{eqnarray}
\mathcal{M}^1_{\;\;\;1}=
\left[\frac{2\kappa^2 \Psi}{2\omega+3  }\, \frac{\partial^2\mathcal{U}}{\partial\Psi^2} \right]_0
\equiv J^{(1)}_{\;\;\;(1)} \,.
\end{eqnarray}
The deviation is simply
\begin{equation}
\Gamma(r) = -\frac{e^{-\sqrt{J}r}}{2 \omega_0+ 3} 
\end{equation}
and from (\ref{gengamma}) it is easy to recognize the familiar result for $\gamma$ \cite{Olmo_Perivolaropoulos,meie2013}
\begin{eqnarray}
\gamma=\frac{2\omega_0+3-e^{-\sqrt{\frac{2\kappa^2\Psi_0}{2\omega_0+3} \left.\frac{\partial^2\mathcal{U}}{\partial\Psi^2}\right|_0 }r}}{2\omega_0+3+e^{-\sqrt{\frac{2\kappa^2\Psi_0}{2\omega_0+3}\left.\frac{\partial^2\mathcal{U}}{\partial\Psi^2}\right|_0}r}} \,.
\end{eqnarray}

\section{Geometric interpretation}
\label{geometric_interpretation_sec}
\label{Sec_4}

\label{section_geometric_interpretation}

In the previous section we found that in MSTG the difference of the effective gravitational constant (\ref{G_eff}) and the PPN parameter $\gamma$ (\ref{gengamma}) from their general relativity values is given by a rather complicated term (\ref{Gamma_def}).
However, when the mass matrix is diagonalizable, i.e.\ it has a complete set of linearly independent eigenvectors this term can be given a neat geometric interpretation. In the current section we show this works in the physically most relevant case when there are no ghosts and the metric on the field space is positive definite, by discussing the mass matrix eigenvalues in section \ref{Sec_4.1} and eigenvectors in section \ref{Sec_4.2}. As an illustrative example this construction is applied to a two field case in the Brans-Dicke like parametrization in section \ref{Sec_4.3}.

\subsection{Eigenvalues of the mass matrix}
\label{Sec_4.1}
The mass matrix $\boldsymbol{\mathcal{M}}$ is given by a product (\ref{mass_matrix}) of the matrix $\boldsymbol{\mathcal{F}}^{-1}$, which is the inverse of the scalar fields' space metric $\mathcal{F}_{\alpha\beta}$,
and the matrix $\boldsymbol{\mathcal{U}}$, proportional to the second partial derivatives of the potential, $\frac{\partial^2 \mathcal{U} }{\partial \Phi^\alpha \partial \Phi^\beta}$, both evaluated at the spatial infinity (i.e.\ at the background values of the scalar fields). Although both $\boldsymbol{\mathcal{F}}^{-1}$ and $\boldsymbol{\mathcal{U}}$ are by construction real symmetric matrices and therefore diagonalizable, their product $\boldsymbol{\mathcal{M}}$ is not automatically so. 


Following Ref.\ \cite{citekey} let us assume that the metric on the space of the scalar fields, $\boldsymbol{\mathcal{F}}$, is positive definite, i.e.\ all the scalars are dynamical and not ghosts. From elementary algebra we know that since $\mathcal{F}_{\alpha\beta}$ is real and symmetric it can be diagonalized by an orthogonal matrix $\boldsymbol{A}$,
\begin{equation}
\label{F_eigenvalues}
(A^T)_{(\theta)}^{\;\;\;\; \epsilon} \, \mathcal{F}_{\epsilon \alpha} A^{\alpha}_{\;\;\; (\gamma)} 
= (\Delta^2)_{(\theta)(\gamma)}
= f^2_{[\gamma]} \delta_{(\theta) (\gamma)} \,,
\end{equation}
where $\boldsymbol{\Delta}^2$ is a diagonal matrix whose entries are nonnegative eigenvalues $f^2_{[\gamma]}$ of $\boldsymbol{\mathcal{F}}$. 
Multiplying Eq.\ (\ref{F_eigenvalues}) from both sides by
$(\Delta^{-1})^{(\gamma)}_{ \phantom{(\gamma)} (\beta) }  \equiv f^{-1}_{[\gamma]} \delta^{(\gamma)}_{ (\beta) } = (\Delta^{-T})_{(\beta)}^{\phantom{(\beta)} (\gamma)}$
normalizes it to
\begin{equation}
\label{F_normalized}
(\Delta^{-T})_{(\zeta)}^{\phantom{(\zeta)}(\theta)} (A^T)_{(\theta)}^{\phantom{(\mu)}\epsilon} \, \mathcal{F}_{\epsilon \alpha} A^{\alpha}_{ \phantom{\alpha} (\gamma)} (\Delta^{-1})^{(\gamma)}_{\phantom{(\gamma)} (\beta) } = \delta_{(\zeta)(\beta)} \,.
\end{equation}
As the matrix
\begin{equation}
\label{Uab_def}
\mathcal{U}_{\epsilon \alpha} = \frac{\kappa^2}{2 \mathcal{F}} \, \frac{\partial^2 \mathcal{U}}{\partial \Phi^\epsilon \partial \Phi^\alpha}
\end{equation}
is symmetric by construction, sandwiching it like Eq.\ (\ref{F_normalized}) also yields a symmetric matrix, where transposing gives
\begin{equation}
\label{Uab_sandwich}
\left( (\Delta^{-T})_{(\zeta)}^{ \phantom{(\zeta)}(\theta)} (A^T)_{(\theta)}^{ \phantom{(\theta)} \epsilon} \, \mathcal{U}_{\epsilon \alpha} A^{\alpha}_{\phantom{\alpha} (\gamma)} (\Delta^{-1})^{(\gamma)}_{ \phantom{(\gamma)} (\beta) } \right)^T
=
(\Delta^{-T})_{(\beta)}^{ \phantom{(\beta)} (\gamma) } (A^T)_{(\gamma)}^{ \phantom{(\gamma)} \alpha} \, \mathcal{U}_{\alpha \epsilon} A^{\epsilon}_{ \phantom{\epsilon} (\theta)} (\Delta^{-1})^{(\theta)}_{ \phantom{(\theta)} (\zeta) } \,.
\end{equation}
Therefore there exists another orthogonal matrix $\boldsymbol{B}$ which diagonalizes the matrix above. Thus for $\boldsymbol{\mathcal{U}}$ we can write
\begin{equation}
\label{U_diagonalized}
(P^T)_{(\eta)}^{\;\;\;\; \epsilon} \, \mathcal{U}_{\epsilon \alpha} P^{\alpha}_{\;\;\; (\delta)} =  m^2_{[\delta]} \delta_{(\eta)(\delta)} \,,
\end{equation}
where the coefficients $m^2_{[\delta]}$ are real and the transformation matrix is
\begin{eqnarray}
\label{P_def}
P^{\alpha}_{\phantom{\alpha} (\delta)} &=& A^{\alpha}_{\phantom{\alpha} (\gamma)} (\Delta^{-1})^{(\gamma)}_{ \phantom{(\gamma)} (\beta) } B^{(\beta)}_{ \phantom{(\beta)} (\delta) } \,,  \\
\label{P^T_def}
(P^T)_{(\eta)}^{\;\;\;\; \epsilon} &=& (B^T)_{(\eta)}^{\phantom{(\eta)}(\zeta)} (\Delta^{-T})_{(\zeta)}^{\phantom{(\eta)}(\theta)} (A^T)_{(\theta)}^{\;\;\;\; \epsilon} \,. 
\end{eqnarray}
The matrix $\boldsymbol{P}$ is normalized with respect to the metric $\boldsymbol{\mathcal{F}}$, 
since from the definitions (\ref{F_eigenvalues}) and (\ref{P_def}), (\ref{P^T_def}) it follows that
\begin{eqnarray}
(P^T)_{(\eta)}^{\;\;\;\; \epsilon} \, \mathcal{F}_{\epsilon \alpha} P^{\alpha}_{\;\;\; (\delta)} &=& \delta_{(\eta)(\delta)} \,, \\
\label{F-1_diagonalized}
(P^{-1})^{(\beta)}_{\;\;\;\; \gamma} \mathcal{F}^{\gamma\zeta} (P^{-T})_\zeta^{\;\;\; (\eta)} &=&  \delta^{(\beta)(\eta)} \,.
\end{eqnarray}
The latter is clear from 
\begin{equation}
(P^{-1})^{(\beta)}_{\;\;\;\; \gamma} \mathcal{F}^{\gamma\zeta} (P^{-T})_\zeta^{\;\;\; (\eta)}
(P^T)_{(\eta)}^{\;\;\;\; \epsilon} \, \mathcal{F}_{\epsilon \alpha} P^{\alpha}_{\;\;\; (\delta)} 
= \delta^{(\beta)}_{(\delta)} \,.
\end{equation}
In principle one could here also say that the entries of $\mathbf{P}$ are vielbeins for the space of the scalar fields. The indices denoted without brackets pertain to a generic basis in the space of the scalar fields and are lowered and raised by the metric $\boldsymbol{\mathcal{F}}$ and its inverse, while the indices denoted in brackets pertain to the orthonormal mass eigenbasis or equivalently to the respective tangent space and are raised and lowered by the flat metric (Kronecker delta).

The matrix $\mathbf{P}$ also diagonalizes the mass matrix $\boldsymbol{\mathcal{M}}$, 
\begin{equation}
\label{J_computation}
J^{(\beta)}_{\;\;\;\; (\delta)} 
= (P^{-1})^{(\beta)}_{\;\;\;\; \gamma} \mathcal{M}^\gamma_{\;\;\; \alpha} P^{\alpha}_{\;\;\; (\delta)} 
= (P^{-1})^{(\beta)}_{\;\;\;\; \gamma} \mathcal{F}^{\gamma\zeta} (P^{-T})_\zeta^{\;\;\; (\eta)} (P^T)_{(\eta)}^{\;\;\;\; \epsilon} \, \mathcal{U}_{\epsilon \alpha} P^{\alpha}_{\;\;\; (\delta)} 
= m^2_{[\delta]} \delta^{(\beta)}_{(\delta)}
 \,,
\end{equation}
due to (\ref{F-1_diagonalized}) and (\ref{U_diagonalized}). 
The mass matrix eigenvalues $m^2_{[\delta]}$ coincide with the eigenvalues of the matrix (\ref{Uab_sandwich}). 
Furthermore, the matrix $\mathcal{U}_{\epsilon\alpha}$ is congruent to the transformed matrix in Eq.~(\ref{U_diagonalized}) and according to Sylvester's law of inertia these matrixes have the same numbers of positive, negative, and zero eigenvalues. Therefore the signs of the mass matrix eigenvalues match the signs of the eigenvalues of the Hessian of the potential.

Let us recall from Sec.~\ref{0th_order_section} that the Minkowski background required the potential $\mathcal{U}$ to vanish and have an extremum in the spatial asymptotics. 
If the potential there has a minimum, or the potential is everywhere nonnegative by construction while there are some flat directions (e.g.\ the potential does not depend on some of the scalar fields), then the eigenvalues of the matrix of second derivatives, $\mathcal{U}_{\epsilon\alpha}$,
as well as the eigenvalues of the mass matrix are all nonnegative.

\subsection{Eigenvectors of the mass matrix}
\label{Sec_4.2}

As the mass matrix diagonalizes, it possesses a full set of linearly independent eigenvectors, $v^\alpha_{\;\;\;(\delta)}$. The components of eigenvectors can be read off from the columns of the similarity matrix,
\begin{equation}
P^\alpha_{\;\;\; (\delta)} = v^\alpha_{\;\;\; (\delta)} \,, 
\qquad
(P^{-1})^{(\beta)}_{\;\;\;\; \gamma} = \delta^{(\beta)(\alpha)} \, v_{(\alpha)}^{\phantom{(\alpha)}\epsilon}  \, \mathcal{F}_{\epsilon \gamma }  \,.
\end{equation}
The eigenvectors are orthonormal, 
\begin{equation}
v_{(\eta)}^{\;\;\;\; \epsilon} \, \mathcal{F}_{\epsilon \alpha} v^{\alpha}_{\;\;\; (\delta)} = (P^T)_{(\eta)}^{\;\;\;\; \epsilon} \, \mathcal{F}_{\epsilon \alpha} P^{\alpha}_{\;\;\; (\delta)} = \delta_{(\eta)(\delta)} \,,
\end{equation}
and by construction satisfy
\begin{equation}
\mathcal{M}^\gamma_{\;\;\; \alpha} v^\alpha_{\;\;\; (\delta)} = m^2_{[\delta]} v^\gamma_{\phantom{\gamma}(\delta)} \,,
\end{equation}
which is obvious from multiplying Eq.\ (\ref{J_computation}) from left by $\boldsymbol{P}$.

As the mass matrix diagonalizes into $\boldsymbol{J}$, taking its square root and exponent are straightforward, and thus
\begin{equation}
E^{(\beta)}_{\;\;\;\; (\delta)} =
\left( e^{-\sqrt{J} \,r} \right)^{(\beta)}_{\;\;\;\; (\delta)} = e^{-m_{[\delta]} r}  \, \delta^{(\beta)}_{(\delta)} \;.
\end{equation}
The deviation term (\ref{Gamma_def_k}) can now be unraveled as
\begin{eqnarray}
\Gamma(r) &=& -\frac{4 \mathcal{F}_0^2}{\kappa^4}  k_\alpha P^\alpha_{\;\;\;(\beta)} 
E^{(\beta)}_{\;\;\;\; (\delta)} (P^{-1})^{(\delta)}_{\;\;\;\;\gamma} k^{\gamma} 
=  -\frac{4 \mathcal{F}_0^2}{\kappa^4} k^\epsilon \, \mathcal{F}_{\epsilon \alpha} \, v^\alpha_{\phantom{\alpha}(\beta)} \; e^{-m_{[\delta]} r}  \, \delta^{(\beta)}_{(\delta)} \; \delta^{(\delta)(\eta)} \, v_{(\eta)}^{\phantom{(\eta)}\zeta}\, \mathcal{F}_{\zeta\gamma } \, k^\gamma 
\nonumber \\
&=&  -\frac{4 \mathcal{F}_0^2}{\kappa^4}  |\boldsymbol{k}|^2 \sum_{\delta} \cos^2 (\vartheta_{(\delta)}) e^{-m_{[\delta]} r} \,,
\label{Gamma_interpretation}
\end{eqnarray}
where the scalar product of the mass matrix eigenvector, $\boldsymbol{v}_{(\delta)}$, and the vector of nonminimal coupling in spatial asymptotics, $\boldsymbol{k}$, has been written in terms of the angle $\vartheta_{(\delta)}$ between them.

The last result informs us that when all scalar fields are massless, the deviation $\Gamma$ in the gravitational constant and PPN parameter $\gamma$ is proportional to the ``strength'' of nonminimal coupling as measured by the length squared of the vector $\boldsymbol{k}$. If the scalars are massive, each mass eigenvalue will give a contribution that reduces the deviation exponentially in spatial distance from the source.
These contributions are weighted according to the angles between the respective eigenvector and the overall vector of nonminimal coupling. For instance a mass eigenvalue whose eigenvector happens to be perpendicular to the vector of nonminimal coupling will not affect the deviation. However, if the vector of nonminimal coupling vanishes, i.e.\ the scalars are minimally coupled, the deviation $\Gamma$ will be zero, irrespective of the masses of the scalars.

\subsection{$N=2$ scalar fields in the Brans-Dicke like parametrization}
\label{Sec_4.3}

In the Brans-Dicke like parametrization (\ref{eq:MSTG:ActionStar_kuusk}) for two scalar fields  $\Phi^1=\phi$, $\Phi^2=\Psi$ and
\begin{equation}
\mathcal{F} = \Psi \,, \qquad
\mathcal{Z}_{\alpha \beta} =
\left(
\begin{array}{cc}
Z(\phi,\Psi) & 0 \\
0 & \frac{\omega(\phi,\Psi)}{\Psi} 
\end{array}
 \right)
\end{equation}
the metric on the space of scalar fields is already diagonal,
\begin{equation}
\mathcal{F}_{\alpha \gamma} =
\left(
\begin{array}{cc}
\frac{Z(\phi,\Psi)}{2\Psi} & 0 \\
0 & \frac{2\omega(\phi,\Psi)+3}{4\Psi^2} 
\end{array}
 \right) \,.
\end{equation}
There are no ghosts as long as $Z \geq 0$ and $2\omega+3 \geq 0$.
The vector of nonminimal coupling has only one nonzero component,
\begin{equation}
\mathcal{K}_\alpha = \left(
\begin{array}{cc}
0 & -\frac{\kappa^2}{4\Psi^2}
\end{array} \right) \,.
\end{equation}
Its square
\begin{equation}
\mathcal{K}_\alpha \mathcal{K}^\alpha = \frac{\kappa^4}{4 \Psi^2 (2\omega+3)}
\end{equation}
tells that the nonminimal coupling disappears in the limit $\frac{1}{2\omega+3}\rightarrow 0$.

The mass matrix
\begin{equation}
\mathcal{M}^{\gamma}_{\;\;\; \alpha} = 
\left[
\begin{array}{cc}
\frac{\kappa^2 }{Z}\frac{\partial^2 \mathcal{U}}{\partial \phi^2} & \frac{\kappa^2 }{Z}\frac{\partial^2 \mathcal{U}}{\partial \phi \, \partial \Psi } \\
\frac{2 \kappa^2 \Psi}{2\omega+3}\frac{\partial^2 \mathcal{U}}{\partial \phi \, \partial \Psi } & \frac{2 \kappa^2 \Psi}{2\omega+3}\frac{\partial^2 \mathcal{U}}{\partial \Psi^2}
\end{array}
 \right]_0 \,
\end{equation}
has eigenvalues
\begin{equation}
m^2_{\pm} = \frac{\kappa^2}{2 Z_0 (2\omega_0+3)} \left((2\omega_0+3)\left. \frac{\partial^2 \mathcal{U}}{\partial \phi^2} \right|_0 + 2 \Psi_0 Z_0 \left. \frac{\partial^2 \mathcal{U} }{\partial \Psi^2 }\right|_0 \pm B \right) \,,
\end{equation}
where
\begin{eqnarray}
B&=&\sqrt{A^2 + 8 (2 \omega_0+3) Z_0 \Psi_0 \left( \left. \frac{\partial^2 \mathcal{U} }{\partial \phi \, \partial \Psi} \right|_0 \right)^2 } \,, \\
A &=& 2 \Psi_0 Z_0 \left. \frac{\partial^2 \mathcal{U} }{\partial \Psi^2 }\right|_0 - (2 \omega_0+3) \left. \frac{\partial^2 \mathcal{U}}{\partial \phi^2}\right|_0 \,.
\end{eqnarray}
One can easily check that both eigenvalues are positive when in the spatial asymptotics the fields are at a minimum of the potential, i.e.\ the Hessian of $\mathcal{U}$ is positive definite there,
\begin{equation}
\left. \frac{\partial^2 \mathcal{U}}{\partial \phi^2} \right|_0 > 0 \,, \qquad
\left. \frac{\partial^2 \mathcal{U}}{\partial \phi^2}\right|_0  \left. \frac{\partial^2 \mathcal{U}}{\partial \Psi^2} \right|_0 >  \left( \left. \frac{\partial^2 \mathcal{U}}{\partial \phi \, \partial \Psi} \right|_0 \right)^2 \,.
\end{equation}
The $\boldsymbol{\mathcal{F}}$-normalized eigenvectors
\begin{equation}
v^\alpha_{\;\; (+)} = \frac{\sqrt{2} \Psi_0 \sqrt{1+\frac{A}{B}}}{\sqrt{2\omega_0+3}} 
\left(
\begin{array}{c}
\frac{2 (2\omega_0+3)}{A+B} \left. \frac{\partial^2 \mathcal{U} }{\partial \phi \, \partial \Psi}\right|_0 \\
1
\end{array} \right) 
\,, \qquad
v^\alpha_{\;\; (-)} = \frac{\sqrt{2} \Psi_0 \sqrt{1-\frac{A}{B}}}{\sqrt{2\omega_0+3}} 
\left(
\begin{array}{c}
\frac{2 (2\omega_0+3)}{A-B} \left. \frac{\partial^2 \mathcal{U} }{\partial \phi \, \partial \Psi} \right|_0 \\
1
\end{array} \right)
\end{equation}
are $\boldsymbol{\mathcal{F}}$-orthogonal and give the columns of the similarity matrix $\boldsymbol{P}$.

Knowing that
\begin{equation}
E^{(\beta)}_{\;\;\;(\delta)}=
\left( \begin{array}{cc}
e^{-m_{+} r} & 0 \\
0 & e^{-m_{-} r}
\end{array} \right)
\end{equation}
we can now calculate from Eq.\ (\ref{Gamma_def_k})
\begin{eqnarray}
\Gamma(r)
&=& -\frac{4 \Psi_0^2}{\kappa^4} \left[ k_\alpha P^\alpha_{\;\;\;(\beta)} 
E^{(\beta)}_{\;\;\;(\delta)} (P^{-1})^{(\delta)}_{\;\;\;\gamma} k^{\gamma} \right] \nonumber \\
&=& - \frac{1}{2\omega_0 + 3} \left( \frac{1}{2}\Big(1+\frac{A}{B} \Big) e^{-m_{+} r} + \frac{1}{2}\Big(1-\frac{A}{B} \Big) e^{-m_{-} r} \right) \,.
\end{eqnarray}
It can be deduced by comparison with Eq.\ (\ref{Gamma_interpretation}) or by direct calculation that the angles between the mass eigenvectors and the vector of nonminimal coupling obey
\begin{equation}
\cos^2 \vartheta_+ = \frac{ (k_\alpha v^\alpha_{\;\;(+)})^2}{|\boldsymbol{k}|^2} = \frac{1}{2}\Big(1+\frac{A}{B} \Big) \,, \qquad
\cos^2 \vartheta_- = \frac{ (k_\alpha v^\alpha_{\;\;(-)})^2}{|\boldsymbol{k}|^2} = \frac{1}{2}\Big(1-\frac{A}{B} \Big) \,.
\end{equation}
Since the mass eigenvectors are orthogonal to each other, $\vartheta_-=\vartheta_+ + \frac{\pi}{2}$, it holds that $\cos^2 \vartheta_+ + \cos^2 \vartheta_- = \cos^2 \vartheta_+ + \sin^2 \vartheta_+ = 1$.

Finally
\begin{equation}
G_{\mathrm{eff}} = \frac{\kappa^2}{8 \pi \Psi_0} \left( 1 + \frac{\cos^2 \vartheta_+ \, e^{-m_+ r} + \cos^2 \vartheta_- \, e^{-m_- r}}{2 \omega_0 + 3} \right)
\end{equation}
and
\begin{equation}
\label{gamma_BD_2fields}
\gamma = \frac{2\omega_0+3 - \cos^2 \vartheta_+ \, e^{-m_+ r} - \cos^2 \vartheta_- \, e^{-m_- r}}{2\omega_0+3 + \cos^2 \vartheta_+ \, e^{-m_+ r} + \cos^2 \vartheta_- \, e^{-m_- r}} \,.
\end{equation}


\section{Observational constraints}
\label{Sec_5}
\label{sec:obs}

Making use of the results derived in the previous two sections, we are now in the position to derive observational constraints on multiscalar-tensor theories of gravity. These will be obtained from the Cassini tracking experiment, whose results we briefly describe in section \ref{Sec_5.1}. In section \ref{Sec_5.2} we then derive constraints on the biscalar theory in the Brans-Dicke like parametrization,  discussed before in section \ref{Sec_4.3}.

\subsection{The Cassini measurement of $\gamma$}
\label{subsec:gammaobs}
\label{Sec_5.1}

Since our result for $\gamma$ generally depends on the interaction distance $r$ between the gravitating source and the test mass acted upon, to get a rough estimate we should turn to an experiment with a clear characteristic interaction distance $r = r_0$.\footnote{As in STG, in a more rigorous analysis one has to integrate over geodesics \cite{Deng:2016moh} and give up the idealization of the point mass \cite{Zhang:2016njn}.} 
The most precise value for $\gamma$ has been obtained from the time delay of radar signals sent between Earth and the Cassini spacecraft on its way to Saturn~\cite{Bertotti:2003rm}. The experiment yielded the value $\gamma - 1 = (2.1 \pm 2.3) \cdot 10^{-5}$ (at $1\sigma$ precision). The radio signals were passing by the Sun at a distance of $1.6$ solar radii or $r_0 \approx 7.44 \cdot 10^{-3}\mathrm{AU}$.

\subsection{Observational constraints for two scalar fields}
\label{Sec_5.2}

In the case of two scalar fields in the Brans-Dicke like parametrization the expression for the PPN parameter $\gamma$ (\ref{gamma_BD_2fields}) involves five quantities which characterize the field configuration in the spatial asymptotic background: the coefficient $\omega_0$ from the kinetic term of the nonminimal scalar, two mass eigenvalues $m_+$, $m_-$, and two angles $\vartheta_+$, $\vartheta_-$ between the mass eigenvectors and the vector of nonminimal coupling in the space of scalar fields. The two angles are related by $\vartheta_+=\vartheta_- - \frac{\pi}{2}$ and the formula (\ref{gamma_BD_2fields}) is symmetric for the interchange of the masses and a reflection across the angle $\frac{\pi}{4}$.

\begin{figure}
\centering
\includegraphics[width=0.32\textwidth]{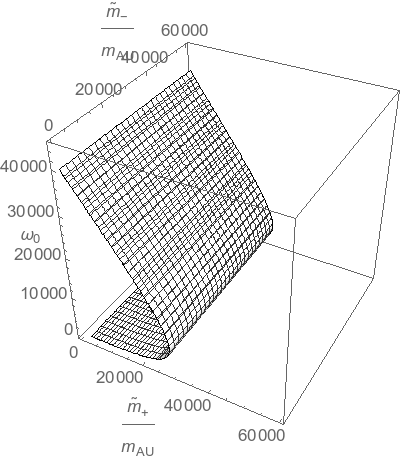}
\includegraphics[width=0.32\textwidth]{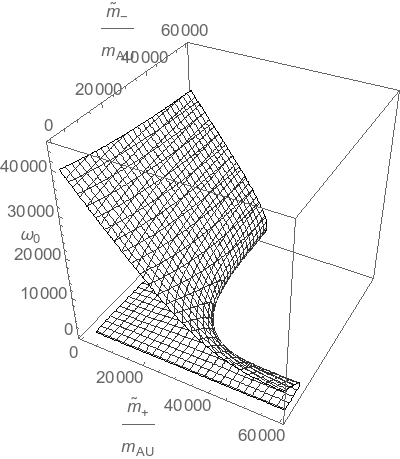}
\includegraphics[width=0.32\textwidth]{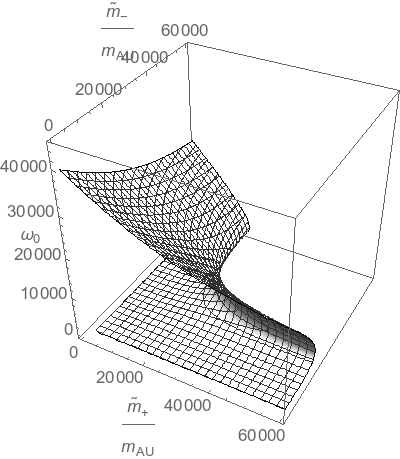}
\caption{Constraints at $2\sigma$ from the Cassini measurement of the PPN parameter $\gamma$ on the rescaled masses of the two scalar fields and the parameter $\omega_0$ for $\vartheta_+=0$ (left), $\vartheta_+=\frac{\pi}{8}$ (middle), and $\vartheta_+=\frac{\pi}{4}$ (right). The allowed region of the parameter space is to the right of the plotted surface.}
\label{gamma_bounds}
\end{figure}

Therefore from the Cassini experimental bounds on the parameter $\gamma$ \cite{Bertotti:2003rm}, taken at the $2\sigma$ confidence level, we can infer rough bounds on the possible values of the theory parameters as plotted on Figure \ref{gamma_bounds}. For better visualization and in order to facilitate comparison with the single field case \cite{meie2013} the horizontal axes depict rescaled masses $\tilde{m}_\pm=\sqrt{2\omega_0+3} \, m_\pm$ normalized by the mass corresponding to the astronomical unit. The vertical axis shows only $\omega_0>-\frac{3}{2}$ since we have assumed the absence of ghosts. The three plots on Fig.\ \ref{gamma_bounds} correspond to the angles $\vartheta_+=0$, $\vartheta_+=\frac{\pi}{8}$, and $\vartheta_+=\frac{\pi}{4}$. The plots for $\vartheta_+=\frac{3\pi}{8}$ and $\vartheta_+=\frac{\pi}{2}$ would be identical to the second and first plot, respectively, with $\tilde{m}_+$ and $\tilde{m}_-$ interchanged.

The allowed parameter region is to the right of the plotted surface. When both fields are massless the bounds disappear in the limit $\omega_0\rightarrow -\frac{3}{2}$ and $\omega_0 \gtrsim 40000$. The existence of masses makes other values of $\omega_0$ also feasible and for sufficiently high masses there are no bounds on $\omega_0$. The graph for a single nonminimal scalar  \cite{meie2013} is identical with a slice of the left plot, since there the eigenvector corresponding to the second mass $m_-$ is perpendicular to the vector of nonminimal coupling and $m_-$ does not have any effect. For arbitrary angles the inclusion of the second massive field reduces the allowed region on one side and extends it on the other side.
It is interesting that at a generic angle $\vartheta$ even a massless field will still have an effect on the result.
This is visible on the middle and right plots, where the bound on $\omega_0$ is much lower than 40000 at the edges of the graph where one mass is zero and another nonzero.

\section{Some specific models}
\label{Sec_6}

To illustrate the formalism developed in Sec.\ \ref{Sec_3} and Sec.\ \ref{Sec_4} let us consider some examples of models formulated as MSTG or equivalent to MSTG.
In particular, we discuss the nonminimally coupled Higgs SU(2) complex doublet in section \ref{Sec_6.1}, hybrid metric-Palatini gravity in section \ref{Sec_6.2}, and linear $\square^{-1}$ and quadratic $\square^{-2}$ nonlocal gravity in sections \ref{Sec_6.3} and \ref{Sec_6.4}.

\subsection{Nonminimally coupled Higgs SU(2) doublet}
\label{Sec_6.1}

Models with the Higgs field nonminimally coupled to curvature are built from the action \cite{Bezrukov:2007ep, Higgs dark energy}
\begin{equation}
\label{S_nmH}
S_{nmH} = \int d^4x \sqrt{-g} \left( \left(\frac{m_p^2}{2} + \xi \mathcal{H}^\dagger \mathcal{H} \right) R - (D_\mu \mathcal{H})^\dagger (D^\mu \mathcal{H}) - \frac{1}{4} F^2 - \frac{\lambda}{4}(\mathcal{H}^\dagger \mathcal{H} - v^2)^2 \right) \,,
\end{equation}
where $D_\mu$ denotes the gauge covariant derivative, and $F$ is the gauge field strength, $\lambda$ is the Higgs self-coupling constant, and $v$ is the Higgs vacuum expectation value. It is convenient to parametrize the Higgs complex doublet by four real scalars as
\begin{equation}
\label{Higgs parametrization}
\mathcal{H} = \frac{1}{\sqrt{2}} 
\left( \begin{array}{c}
\phi_1 e^{i \theta_1} \\
\phi_2 e^{i \theta_2}
\end{array}
\right) \,.
\end{equation}
We assume the gauge fields do not play a role in the typical scales of e.g.\ the Solar System and neglect them. 
By equating the Planck mass $m_p=\frac{1}{\kappa}$, and making the scalars dimensionless, $\phi_1 =  \frac{\Phi^1}{\kappa}$, $\phi_2 = \frac{\Phi^2}{\kappa}$, $\theta_1 = \Phi^3$, $\theta_2 = \Phi^4$ we can write the action (\ref{S_nmH}) in the form of the general MSTG action (\ref{eq:MSTG:Action_kuusk}), with
\begin{equation}
\mathcal{F} = 1 + \xi \left( (\Phi^1)^2 + (\Phi^2)^2 \right) \,, \quad
\mathcal{Z}_{\alpha \beta} = \mathrm{diag} \left(1, 1, (\Phi^1)^2, (\Phi^2)^2 \right) \,, \quad
\mathcal{U} = \frac{\lambda}{4 \kappa^4} \left( \frac{(\Phi^1)^2}{2}  + \frac{(\Phi^2)^2}{2}  - \kappa^2 v^2 \right)^2 \,.
\end{equation}
The metric of the space of scalar fields (\ref{F_ab}),
\begin{equation}
\label{Higgs F_ab}
\mathcal{F}_{\alpha \beta} = \left( 
\begin{array}{cccc}
\frac{\mathcal{F}+6 \xi^2 (\Phi^1)^2}{2\mathcal{F}^2} & \frac{3 \xi^2 \Phi^1 \Phi^2}{\mathcal{F}^2} & 0 & 0 \\
\frac{3 \xi^2 \Phi^1 \Phi^2}{\mathcal{F}^2} & \frac{\mathcal{F}+6 \xi^2 (\Phi^2)^2}{2\mathcal{F}^2} & 0 & 0 \\
0 & 0 & \frac{ (\Phi^1)^2}{2 \mathcal{F}} & 0 \\
0 & 0 & 0& \frac{ (\Phi^2)^2}{2 \mathcal{F}} 
\end{array} \right)
\end{equation}
is positive definite (and there are no ghosts) as long as $\mathcal{F}>0$. The vector of nonminimal coupling (\ref{k def}),
\begin{equation}
\mathcal{K}_\alpha = \left( \begin{array}{cccc}
-\frac{\kappa^2 \xi \Phi^1}{2 \mathcal{F}^2} & -\frac{\kappa^2 \xi \Phi^2}{2 \mathcal{F}^2} &0 &0 
\end{array} \right) \,,
\end{equation}
tells that in the parametrization (\ref{Higgs parametrization}) only two of the four Higgs components have a direct nonminimal coupling to curvature, but the other two components are still indirectly involved via the metric (\ref{Higgs F_ab}).

For the PPN setup in the spatial background the Higgs field must reside at the minimum of the potential,
\begin{equation}
(\Phi_0^1)^2 + (\Phi_0^2)^2 = 2 \kappa^2 v^2 \,,
\end{equation}
to satisfy the conditions (\ref{U 0}) and (\ref{U deriv 0}).
The mass matrix (\ref{mass_matrix}) can be found by a straightforward computation, it has only one nonzero eigenvalue,
\begin{equation}
m^2_{[1]} = m^2_{H} = \frac{\lambda v^2 (1 + 2 \xi \kappa^2 v^2)}{1+2\xi \kappa^2 v^2 + 12 \xi^2 \kappa^2 v^2} \,,
\end{equation}
while the other eigenvalues $m^2_{[2]}=m^2_{[3]}=m^2_{[4]}=0$. We see that by the virtue of nonminmal coupling the usual Higgs mass expression gets a $\xi$-dependent correction. However, since the Higgs vacuum expectation value $v$ is many orders of magnitude smaller than the Planck mass ($\kappa^2 v^2 \sim 10^{-34}$) this correction is really tiny for typical values of the nonminimal coupling constant required by the Higgs inflation ($\xi \sim 10^4$ \cite{Bezrukov:2007ep, Higgs dark energy}). 

There exists a full set of eigenvectors, orthonormal with respect to the metric (\ref{Higgs F_ab}), encoded in the columns of the similarity matrix 
\begin{equation}
\mathcal{P}^{\alpha}_{\;\;\; (\beta)} = \sqrt{1+2\xi\kappa^2 v^2} \left( 
\begin{array}{rrrr}
\frac{\sqrt{1 + 2 \xi \kappa^2 v^2}}{\sqrt{1+2\xi \kappa^2 v^2 + 12 \xi^2 \kappa^2 v^2}} \frac{\Phi^1_0}{\kappa v} & -\frac{\Phi^2_0}{\kappa v} & 0 & 0 \\
 \frac{\sqrt{1 + 2 \xi \kappa^2 v^2}}{\sqrt{1+2\xi \kappa^2 v^2 + 12 \xi^2 \kappa^2 v^2}} \frac{\Phi^2_0}{\kappa v} &  \frac{\Phi^1_0}{\kappa v} & 0 & 0 \\
0 & 0 & \frac{\sqrt{2}}{\Phi^1_0} & 0 \\
0 & 0 & 0&  \frac{\sqrt{2}}{\Phi^2_0}
\end{array} \right)
\end{equation}
Inserting these results into Eq.\ (\ref{Gamma_def_k}), the deviation from general relativity is found to be
\begin{equation}
\Gamma = - \frac{4 \xi^2 \kappa^2 v^2 \, e^{-m_H r}}{1 + 2 \xi \kappa^2 v^2 + 12 \xi^2 \kappa^2 v^2} \,.
\end{equation}
Therefore the effective gravitational constant (\ref{G_eff}) is
\begin{eqnarray}
G_{\mathrm{eff}} &=& \frac{\kappa^2}{8 \pi (1+ 2\xi \kappa^2 v^2)} \left(1 + \frac{4 \xi^2 \kappa^2 v^2 \, e^{-m_H r}}{1 + 2 \xi \kappa^2 v^2 + 12 \xi^2 \kappa^2 v^2} \right) 
\end{eqnarray}
and the PPN parameter $\gamma$ (\ref{gengamma}) is given by 
\begin{eqnarray}
\gamma &=& \frac{1 + 2 \xi \kappa^2 v^2 + 12 \xi^2 \kappa^2 v^2 - 4 \xi^2 \kappa^2 v^2 \, e^{-m_H r}}{1 + 2 \xi \kappa^2 v^2 + 12 \xi^2 \kappa^2 v^2 + 4 \xi^2 \kappa^2 v^2 \, e^{-m_H r}} \,.
\end{eqnarray}
Since for the Standard Model Higgs and the Cassini experiment characteristic distance the combination $m_Hr \sim 10^{26}$, the predicted value of $\gamma$ is well within the observational bounds. This concurs with the estimate about the single field Higgs monopole configuration \cite{Higgs_monopole}.

\subsection{General hybrid metric-Palatini}
\label{Sec_6.2}

General hybrid metric-Palatini gravity combines curvature $R$ computed from the metric and curvature $\mathcal{R}$ computed from independent connection into a single action
\cite{general hybrid metric-Palatini}
\begin{equation}
S_{ghmP} = \frac{1}{2\kappa^2} \int d^4x \sqrt{-g} f(R, \mathcal{R}) \,.
\end{equation}
By introducing two scalars it is possible to rewrite the action as \cite{general hybrid metric-Palatini}
\begin{equation}
S_{ghmP} = \frac{1}{2\kappa^2} \int d^4x \sqrt{-g} \left[ \phi R - \frac{3}{2\xi} g^{\mu \nu} \partial_\mu \xi \partial_\nu \xi - W(\phi,\xi) \right]  \,,
\end{equation}
where the potential $W$ encodes the original function $f$. The latter action is in the MSTG form, matching Eq.\ (\ref{eq:MSTG:Action_kuusk}) with the identifications $\phi=\Phi^1$, $\xi=\Phi^2$, and
\begin{equation}
\mathcal{F} = \Phi^1 \,, \qquad \mathcal{Z}_{22} = \frac{3}{2 \Phi^2} \,, \qquad \mathcal{U}=\frac{1}{2\kappa^2} W(\Phi^1,\Phi^2).
\end{equation}
The eigenvalues of the metric of the space of scalar fields (\ref{F_ab}),
\begin{equation}
\label{F_ab ghmP}
\mathcal{F}_{\alpha \beta} = 
\left( 
\begin{array}{cc}
\frac{3}{4 (\Phi^1)^2} & 0 \\
0 & \frac{3}{4 \Phi^1 \Phi^2}
\end{array}
\right)
\end{equation}
can be read off from the diagonal, there are no ghosts provided that $\Phi^1$ and $\Phi^2$ are both positive. The vector of nonminimal coupling (\ref{k def}),
\begin{equation}
\label{K ghmP}
\mathcal{K}_\alpha = \left( \begin{array}{cc}
-\frac{\kappa^2}{4 (\Phi^1)^2} & 0 
\end{array} \right) \,,
\end{equation}
tells that in this parametrization only one of the scalar fields has a direct nonminimal coupling, but the other is indirectly involved via the potential and the metric (\ref{F_ab ghmP}).
It is not possible to write out the Jordan form of the mass matrix for a generic potential (there are several distinct possibilities). Therefore let us look at the models previously considered in the literature \cite{general hybrid metric-Palatini}. 

Model 1 is characterized by
\begin{equation}
W = W_0 e^{\frac{-\lambda \Phi^1}{\sqrt{6}}} \,.
\end{equation}
The conditions for the Minkowski background (\ref{U 0}), (\ref{U deriv 0}) are satisfied in the limit $\Phi^1 \rightarrow \infty$ and the mass matrix (\ref{mass_matrix}) vanishes. Due to the nontrivial vector of nonminimal coupling (\ref{K ghmP}) the factor (\ref{Gamma_def_k}) is $\Gamma=-\frac{1}{3}$, but from Eqs.\ (\ref{G_eff}), (\ref{gengamma})
\begin{equation}
G_{\mathrm{eff}} \rightarrow 0 \,, \qquad \gamma=\frac{1}{2} \,
\end{equation} 
and the model is not viable.

Model 2 is characterized by
\begin{equation}
W = W_0 (\Phi^2)^\lambda e^{-\frac{\lambda \Phi^1}{\sqrt{6}}} \,.
\end{equation}
For $\lambda>1$ the conditions for the Minkowski background (\ref{U 0}), (\ref{U deriv 0}) can now be satisfied also by $\Phi^2_0 = 0$. However the mass matrix (\ref{mass_matrix}) vanishes again and result is the same as for model 1.

\subsection{Linear ($\Box^{-1}$) nonlocal gravity}
\label{Sec_6.3}

The simplest example to build a nonlocal gravity is to include to the action an inverse of the d'Alembertian operator acting on the Ricci scalar \cite{Deser:2007jk},
\begin{equation}
\label{nonlocal_1_action}
S_{NL^{-1}} = \frac{1}{2 \kappa^2} \int d^4x \sqrt{-g} \, R \left(1 + f \left( \Box^{-1}R \right) \right) \,.
\end{equation}
Adding a suitable Lagrange multiplier and performing an integration by parts to replace the $\Box$-term allows to write this action in the MSTG form as \cite{nonlocal}
\begin{equation}
S_{NL^{-1}} = \frac{1}{2 \kappa^2} \int d^4x \sqrt{-g} \,  \left( (1 + f(\Phi^1) + \Phi^2)R + g^{\mu \nu} \partial_\mu \Phi^1 \partial_\nu \Phi^2  \right) \,.
\end{equation}
Here in our notation $\mathcal{F}=1+f(\Phi^1)+\Phi^2$ and the metric on the space of scalar fields (\ref{F_ab}),
\begin{equation}
\mathcal{F}_{\alpha \beta} = 
\frac{1}{4 \mathcal{F}^2}
\left( \begin{array}{cc}
3 \left( \frac{\partial f}{\partial \Phi^1} \right)^2 & 
-\mathcal{F} + 3 \left( \frac{\partial f}{\partial \Phi^1} \right) \\
-\mathcal{F} + 3 \left( \frac{\partial f}{\partial \Phi^1} \right) & 3
\end{array} \right) \,,
\end{equation}
is positive definite (i.e.\ there are no ghosts) if
\begin{equation}
\frac{6}{\mathcal{F}} \frac{\partial f}{\partial \Phi^1} > 1 \,.
\end{equation}
This inequality matches exactly the result obtained via multiple intermediate redefinitions of the scalar fields and transforming into the Einstein frame \cite{no ghosts in linear nonlocal gravity,sasaki nonlocal}. 
In fact, the action (\ref{nonlocal_1_action}) provides the only member in the family of nonlocal gravities which can be free of ghosts in the MSTG representation \cite{sasaki nonlocal}. There is no potential and all the fields are massless, hence deviations from general relativity (\ref{Gamma_def_k}) come from the nonminimal coupling only,
\begin{equation}
\Gamma = - \frac{4 \mathcal{F}_0^2}{\kappa^4} k_\alpha k^\alpha = \left[ \frac{2 \frac{\partial f}{\partial \Phi^1 }}{(\mathcal{F}- 6 \frac{\partial f}{\partial \Phi^1 })} \right]_0 
\end{equation}
From the general formulas (\ref{G_eff}) and (\ref{gengamma}) we can now read off 
\begin{eqnarray}
G_{\mathrm{eff}} &=& \frac{\kappa^2}{8 \pi \mathcal{F}_0} \left[ \frac{\mathcal{F}-8 \frac{\partial f}{\partial \Phi^1 } }{ \mathcal{F}-6 \frac{\partial f}{\partial \Phi^1 } } \right]_0  \\
\gamma &=& \left[ \frac{\mathcal{F}-4 \frac{\partial f}{\partial \Phi^1 } }{ \mathcal{F}-8 \frac{\partial f}{\partial \Phi^1 }  } \right]_0 \,,
\end{eqnarray} 
which is in complete agreement with the earlier calculation using biscalar representation \cite{Koivisto:2008dh} and a subsequent direct calculation \cite{Conroy:2014eja}. Experimental bounds on $\gamma$ can be invoked to constrain the possible forms of the function $f$. Yet, the model has been already disfavored by the cosmological data \cite{Dodelson:2013sma}. 

\subsection{Quadratic ($\Box^{-2}$) nonlocal gravity}
\label{Sec_6.4}

A nonlocal gravity model more viable cosmologically is provided by including to the action the inverse-squared  d'Alembertian operator acting on the Ricci scalar \cite{Maggiore:2014sia},
\begin{equation}
\label{S_NL2}
S_{NL^{-2}_-} = \frac{1}{16 \pi G} \int d^4x \sqrt{-g} \left[ R - \frac{m^2}{6} R \, \Box^{-2} R \right] \,,
\end{equation}
where $m$ is a parameter with mass dimension.
By introducing two scalars $U=-\Box^{-1}R$, $S=-\Box^{-1}U$, it is possible to write the action (\ref{S_NL2}) as \cite{Maggiore:2014sia}
\begin{equation}
\label{S_NL2_scalar}
S_{NL^{-2}} = \frac{1}{16 \pi G} \int d^4x \sqrt{-g} \left[ \left( 1 - \frac{m^2}{6} S \right) R - \xi_1 \left( \Box U + R \right) - \xi_2 \left( \Box S + U \right) \right]
\end{equation}
where $\xi_1$, $\xi_2$ are Lagrange multipliers. 
By adopting the identifications $16 \pi G=2\kappa^2$, $\frac{m^2}{6}=\frac{\mu}{2\kappa^2}$ and in terms of the dimensionless scalar fields, $U=\Phi^1$, $S=2\kappa^2 \Phi^2$, $\xi_1=\Phi^3$, $\xi_2=\frac{1}{2\kappa^2} \Phi^4$, after integration by parts the action (\ref{S_NL2_scalar}) takes the form of the generic MSTG action (\ref{eq:MSTG:Action_kuusk}), with
\begin{equation}
\mathcal{F}=1-\mu \Phi^2 - \Phi^3 \,, \qquad
\mathcal{Z}_{13} = \mathcal{Z}_{31} = \mathcal{Z}_{24} = \mathcal{Z}_{42} = -\frac{1}{2} \,, \qquad
\mathcal{U} = \frac{1}{4 \kappa^4} \Phi^1 \Phi^4 \,.
\end{equation}
The metric of the space of scalar fields (\ref{F_ab}),
\begin{equation}
\label{F_ab NL2}
\mathcal{F}_{\alpha \beta} = \left( \begin{array}{cccc}
0 & 0 & -\frac{1}{4 \mathcal{F}} & 0 \\
0 & \frac{3 \mu^2}{4 \mathcal{F}^2} & \frac{3 \mu}{4 \mathcal{F}^2} &  -\frac{1}{4 \mathcal{F}} \\
-\frac{1}{4 \mathcal{F}} & \frac{3 \mu}{4 \mathcal{F}^2} & \frac{3}{4 \mathcal{F}^2} & 0 \\
0 & -\frac{1}{4 \mathcal{F}} & 0 & 0 
\end{array} \right) \,,
\end{equation}
has eigenvalues
\begin{equation}
\label{F eigenv NL2}
f_{[1,2]} = \pm \frac{1}{4 \mathcal{F}} \,, \qquad
f_{[3,4]} = \frac{3(1+\mu^2) \pm \sqrt{9(1+\mu^2)^2+4\mathcal{F}^2} }{8\mathcal{F}^2}
\end{equation}
which tell that two of the four scalars are actually ghost degrees of freedom. This is in accordance with the observation in Ref.\ \cite{sasaki nonlocal}, however the case with the original theory (\ref{S_NL2_scalar}) is more subtle \cite{Maggiore:2014sia,Maggiore:2016gpx}. The vector of nonminimal coupling (\ref{k def}),
\begin{equation}
\label{Ka_NL2}
\mathcal{K}_\alpha = \left( \begin{array}{cccc}
0 & \frac{\kappa^2 \mu}{4 \mathcal{F}^2} & \frac{\kappa^2}{4 \mathcal{F}^2} & 0
\end{array} \right) \,, 
\end{equation}
has zero length,
\begin{equation}
\mathcal{F}^{\alpha \beta} \mathcal{K}_\alpha \mathcal{K}_\beta = 0 \,.
\end{equation}
The latter property does not mean that the scalars are minimally coupled, but occurs because the metric (\ref{F_ab NL2}) is not positive definite. Such situation is actually rather reasonable. Namely, on the one hand the action (\ref{S_NL2_scalar}) reduces to general relativity with minimally coupled scalars in the $m\rightarrow 0$ limit where the vector of nonminimal coupling should vanish. On the other hand the third component of the vector of nonminimal coupling (\ref{Ka_NL2}) is independent of the mass scale $m$ and is unaffected by this limiting procedure. Therefore for the general relativity limit to exist, the vector (\ref{Ka_NL2}) must be of zero length as measured by the metric (\ref{F_ab NL2}) already for arbitrary value of $m$.

The Minkowski background requires
\begin{equation}
\Phi^1_0 = \Phi^4_0 = 0 \,
\end{equation}
and we assume the values $\Phi^2_0$, $\Phi^3_0$ get fixed by some mechanism. (It is an open issue how much the cosmological evolution of the scalar $S$ could affect local experiments in the Solar System \cite{Barreira:2014kra,Dirian:2016puz}.)
As the Hessian of the potential $\mathcal{U}$ has one positive and one negative nonzero eigenvalue, the mass matrix (\ref{mass_matrix}) 
\begin{equation}
\mathcal{M}^{\gamma}_{\;\;\; \alpha} = 
\left( 
\begin{array}{cccc}
-\frac{3 \mu}{2\kappa^2 \mathcal{F}_0} & 0 & 0 & -\frac{3 }{2\kappa^2 \mathcal{F}_0} \\
-\frac{1}{2\kappa^2} & 0 & 0 & 0 \\
0 & 0 & 0 & -\frac{1}{2\kappa^2} \\
-\frac{3 \mu^2}{2\kappa^2 \mathcal{F}_0} & 0 & 0 & -\frac{3 \mu}{2\kappa^2 \mathcal{F}_0} 
\end{array}
   \right)
\end{equation}
is not diagonalizable, but admits a Jordan form (cf.\ Appendix \ref{App_C})
\begin{equation}
\label{J NL2}
J^{(\beta)}_{\;\;\; (\delta)} = 
\left( 
\begin{array}{cccc}
0 & \mathfrak{m}^2 & 0 & 0 \\
0 & 0 & 0 & 0 \\
0 & 0 & m^2_{[3]} & 0 \\
0 & 0 & 0 & 0 
\end{array}
   \right) \,,
\end{equation}
where the only nonzero mass eigenvalue is 
\begin{equation}
m^2_{[3]} = - \frac{3 \mu}{\kappa^2 \mathcal{F}_0} = -\frac{m^2}{\mathcal{F}_0} \,,
\end{equation}
while $\mathfrak{m}$ is an arbitrary constant of mass dimension, which cancels out later since the similarity matrix is given by
\begin{equation}
\label{P NL2}
P^{\alpha}_{\;\;\; (\delta)} = 
\left( 
\begin{array}{cccc}
0 & \frac{\mathfrak{m}^2}{2} & \frac{1}{2} & 0 \\
-\frac{1}{4 \kappa^2} & \frac{\mu+4}{4 \mu \kappa^2 m^2_{[3]}} & -\frac{1}{4 \kappa^2 m^2_{[3]}} & \frac{1}{\mu \kappa^2 m^2_{[3]}} \\
\frac{\mu}{4 \kappa^2} & \frac{\mu}{4 \kappa^2 m^2_{[3]}} & -\frac{\mu}{4 \kappa^2 m^2_{[3]}} & 0 \\
0 & -\frac{\mu \mathfrak{m}^2}{2} & \frac{\mu}{2} & 0 
\end{array}
   \right) \,.
\end{equation}
The Jordan matrix (\ref{J NL2}) does not have a square root, but the solutions for the scalar fields in the mass basis can be written out as
\begin{eqnarray}
\PPh{2}^{(\beta)} = -\frac{M_0}{4\pi r} E^{(\beta)}_{\;\;\;(\gamma)} \, \hat{k}^{(\gamma)}\,,
\end{eqnarray} 
where
\begin{equation}
E^{(\beta)}_{\;\;\; (\gamma)} = 
\left( 
\begin{array}{cccc}
1 & \frac{\mathfrak{m}^2 r^2}{2} & 0 & 0 \\
0 & 1 & 0 & 0 \\
0 & 0 & c(r) & 0 \\
0 & 0 & 0 & 1 
\end{array}
   \right) \,.
\end{equation}
The oscillating dependence on the distance,
\begin{equation}
\label{c_NL2}
\PPh{2}^{(3)} = 
-\frac{M_0 \hat{k}^{(3)}}{4\pi r} 
c(r)=\frac{c_1 \cos (\sqrt{|m_{[3]}^2|} r) + c_2 \sin (\sqrt{|m_{[3]}^2|} r)}{r} \,,
\end{equation}
arises because $m^2_{[3]}<0$. Here the integration constant $c_1=-\frac{M_0 \hat{k}^{(3)}}{4\pi}$, but $c_2$ remains undetermined, see Appendix \ref{App_B_neg_eigenvalue} for the details.
Now the deviation from the general relativity value (\ref{Gamma_def_k}) is
\begin{eqnarray}
\Gamma(r) &=& -\frac{4 \mathcal{F}_0^2}{\kappa^4} \left[ k_\alpha P^\alpha_{\;\;\;(\beta)} 
E^{(\beta)}_{\;\;\;(\delta)} (P^{-1})^{(\delta)}_{\;\;\;\gamma} k^{\gamma} \right] 
= \frac{\mu (1-c(r))}{\kappa^2 m^2_{[3]} \mathcal{F}_0} = -\frac{1}{3} \left(1 - c(r) \right)
\end{eqnarray}
and the final formulas (\ref{G_eff}), (\ref{gengamma}) yield
\begin{eqnarray}
\label{Geff_NL2}
G_{\mathrm{eff}} &=& \frac{\kappa^2}{8 \pi \mathcal{F}_0} \left(1+\frac{(1-c(r))}{3} \right) \,, \\
\gamma &=& \frac{2+c(r)}{4-c(r)} \,,
\end{eqnarray} 
This result can be compared to the Newtonian limit found in Ref.\ \cite{Kehagias:2014sda} for the intermediate distances much larger than the Schwarzschild radius but much smaller than the scale $m^{-1}$ which is assumed to be comparable to Hubble scale playing a role in cosmology. By a manysided investigation the authors of Ref.\ \cite{Kehagias:2014sda} are able to determine the integration constants. The effective gravitational constant that can be read off from Eqs.\ (4.30), (4.31) in Ref.\ \cite{Kehagias:2014sda} matches our result (\ref{Geff_NL2}) when the integration constant $c_2$ in Eq.\ (\ref{c_NL2}) gets fixed to zero. In that case, for sufficiently small values of the combination $mr$, also the post-Newtonian parameter $\gamma$ will be close to unity and satisfy the observations.

If we change the sign of the nonlocal term,
\begin{equation}
\label{S_NL2+}
S_{NL^{-2}_+} = \frac{1}{16 \pi G} \int d^4x \sqrt{-g} \left[ R + \frac{m^2}{6} R \, \Box^{-2} R \right] \,,
\end{equation}
the MSTG form of the action remains the same, except
\begin{equation}
\mathcal{F}=1+\mu \Phi^2 - \Phi^3 \,.
\end{equation}
The eigenvalues of the metric on the scalar fields are still given by  Eq.\ (\ref{F eigenv NL2}) and two of the four scalars are ghosts. However the only nonzero mass matrix eigenvalue is now positive,
\begin{equation}
m^2_{[3]} = \frac{3 \mu}{\kappa^2 \mathcal{F}_0} = \frac{m^2}{\mathcal{F}_0}\,
\end{equation}
and the distance dependence is exponential. This leads to
\begin{eqnarray}
\Gamma(r) &=&  -\frac{1-e^{-m_{[3]}r}}{3} \,,
\\
G_{\mathrm{eff}} &=& \frac{\kappa^2}{8 \pi \mathcal{F}_0} \left(1+\frac{1-e^{-m_{[3]}r}}{3} \right) \,, \\
\gamma &=& \frac{2+e^{-m_{[3]}r}}{4-e^{-m_{[3]}r}} \,,
\end{eqnarray} 
in agreement with the direct calculation proceeding from the original action in Ref.\ \cite{Conroy:2014eja}.

\section{Summary and outlook}
\label{Sec_7}

In this paper we considered a generic multiscalar-tensor gravity (MSTG) with arbitrary coupling functions and potential (but no derivative couplings) in the Jordan frame, and computed the post-Newtonian parameter $\gamma$ using a point mass as a source. 
In the single field case the result reproduces the earlier study \cite{meie2013}, where a massive nonminimal scalar is known to modify the effective gravitational constant $G_{\mathrm{eff}}$ and the PPN parameter $\gamma$ by a correction which falls off exponentially in distance. 
The same effect persists in the multiscalar case, while  $G_{\mathrm{eff}}$ (\ref{G_eff}) and $\gamma$ (\ref{gengamma})  now depend in an intricate way not only on the masses but also on the alignment of the fields in the field space (\ref{Gamma_def}). 

To describe the geometry of the field space we found it useful to introduce a metric (\ref{F_ab}) and a vector (\ref{k def}) which describes the nonminimal coupling of the scalars to gravity (spacetime curvature). 
These objects transform covariantly under the reparametrization of the scalars, i.e.\ a change of the coordinates and the corresponding local basis in the field space. 
The PPN calculation lead to the mass matrix (\ref{mass_matrix}) whose eigenvectors (or generalized eigenvectors) form a special basis in the field space. 
A nice interpretation can be given if the field space metric is positive definite (there are no ghosts among the scalars), as in Eq.\ (\ref{Gamma_interpretation}) each massive field gives a contribution depending on the angle between the respective mass eigenvector and the overall vector of nonminimal coupling. 

The situation is perhaps easier to grasp in the case of just two fields in the Brans-Dicke like parametrization where the  formula (\ref{gamma_BD_2fields}) for $\gamma$ can be used to plot constraints on the theory parameters from the Cassini tracking  experiment, see Fig.\ \ref{gamma_bounds}. As expected, for massless scalars the bounds on the asymptotic value of the Brans-Dicke $\omega$ are high, while sufficiently large masses remove any bound on $\omega$. A very interesting scenario would arise when one massless scalar is accompanied by a rather massive scalar, since in that case the experimental constraints on $\omega$ will be also greatly reduced compared to a single massless nonminimal scalar (provided the  alignment in the field space is sufficiently favorable).

Our results are very general and can be utilized to test the viability of a multitude of models, either originally formulated as MSTG or shown to be equivalent to MSTG. As an illustration of the formalism we considered four relevant examples: nonminimally coupled Higgs SU(2) complex doublet, general hybrid metric-Palatini gravity, linear ($\Box^{-1}$) nonlocal gravity, and quadratic ($\Box^{-2}$) nonlocal gravity. In the cases where an earlier PPN result was available in the literature for these examples, it agreed with the application of our formulas for that specific model.  

The computations of our paper were carried out in the Jordan frame. However, in various contexts and for several applications it is useful to focus upon the Einstein frame instead. Therefore it remains a task for future to give the results also in the Einstein frame, or even better, rephrase them in terms of the formalism of invariants \cite{meie2015} which facilitates their easy implementation in any frame and parametrization. Our insights to the role of the geometry on the field space are most likely just a first glimpse, and the formalism of invariants seems a fruitful tool to clarify these issues as well. 

The estimation of the numerical bounds on the model parameters assuming a characteristic distance from a point mass can be a rather crude approximation for an actual astrophysical experiment. For a more realistic situation one should integrate over geodesics and invoke an extended source, as has been done for a single nonminimally coupled field case \cite{Deng:2016moh, Zhang:2016njn}. Finally, to fully test the MSTG models the PPN weak field arena must be complemented by research on the strong field regime as well \cite{Berti:2015itd}.

\section*{Acknowledgments}

The authors were supported by the Estonian Ministry for Education and Science Institutional Research Support Project IUT02-27 and Startup Research Grant PUT790, as well as by the European Regional Development Fund through the Center of Excellence TK133. The authors thank Sergey Odintsov and Kevin A.\ S.\ Croker for useful discussions.

\appendix

\section{Diagonalization of the ``mass matrix''}
\label{mass_matrix_eigenvalues}
\label{App_A}

In Sec.\ \ref{Sec_4} we showed that when the metric (\ref{F_ab}) on the space of scalar fields is positive definite (there are no ghosts) the mass matrix (\ref{mass_matrix}) is diagonalizable, which simplifies the calculation and interpretation of the results. In this Appendix we outline some other cases when the the mass matrix also admits a diagonal form, drawing on Ref.\ \cite{citekey}.

For instance if the matrix $\boldsymbol{\mathcal{U}}$ (proportional to the Hessian of the potential, Eq.\ (\ref{Uab_def})) is positive definite, then by a construction similar to Sec.\ \ref{Sec_4.1} we can find a matrix $\boldsymbol{C}$ such that
\begin{equation}
\label{F_eigenvalues_U}
(C^T)_{(\theta)}^{\;\;\;\; \epsilon} \, \mathcal{U}_{\epsilon \alpha} C^{\alpha}_{\;\;\; (\gamma)} 
= \delta_{(\theta) (\gamma)} \,
\end{equation}
and simultaneously
\begin{equation}
\label{F_eigenvalues_UF}
(C^T)_{(\theta)}^{\;\;\;\; \epsilon} \, \mathcal{F}_{\epsilon \alpha} C^{\alpha}_{\;\;\; (\gamma)} 
= \frac{1}{m^2_{[\gamma]}} \delta_{(\theta) (\gamma)} \,.
\end{equation} 
Here the coefficients $\frac{1}{m^2_{[\gamma]}}$ may be also negative. One can normalize the metric $\boldsymbol{\mathcal{F}}$ by introducing a positive diagonal matrix
\begin{equation}
(\bar{\Delta}^{2})_{(\theta) (\gamma)} = \frac{1}{|m^2_{[\gamma]}|} \delta_{(\theta) (\gamma)} \,. 
\end{equation}
Let us define $\mathbf{P} = \mathbf{C} \bar{\Delta}^{-1}$. Then
\begin{equation}
(P^T)_{(\eta)}^{\;\;\;\; \epsilon} \, \mathcal{F}_{\epsilon \alpha} P^{\alpha}_{\;\;\; (\delta)} = (\mathrm{sign}\, m^2_{[\delta]}) \delta_{(\eta)(\delta)}
\end{equation}
which is a diagonal matrix with $+1$ and $-1$ entries depending on the signs of the coefficients $m^2_{[\delta]}$. Finally
\begin{equation}
\label{J_computation_U}
J^{(\beta)}_{\;\;\;\; (\delta)} 
= (P^{-1})^{(\beta)}_{\;\;\;\; \gamma} \mathcal{M}^\gamma_{\;\;\; \alpha} P^{\alpha}_{\;\;\; (\delta)} 
= (P^{-1})^{(\beta)}_{\;\;\;\; \gamma} \mathcal{F}^{\gamma\varepsilon} (P^{-T})_\varepsilon^{\;\;\; (\eta)} (P^T)_{(\eta)}^{\;\;\;\; \epsilon} \, \mathcal{U}_{\epsilon \alpha} P^{\alpha}_{\;\;\; (\delta)} 
= m^2_{[\delta]} \delta^{(\beta)}_{(\delta)}
 \,.
\end{equation}
Therefore the Jordan normal form of the mass matrix $\boldsymbol{\mathcal{M}}$ is again diagonal.
The mass matrix eigenvectors are still orthonormal w.r.t.\ the metric $\boldsymbol{\mathcal{F}}$ and form a complete set.
The components $P^{\alpha}_{\;\;\; (\delta)}$ can be again interpreted as vielbeins on the space of scalar fields, but now with appropriate possibly pseudo-euclidean signature. 
Let us also point out that nothing in the previous construction changes if not all of the eigenvalues are distinct.

So far we have established by an explicit construction that the mass matrix is diagonalizable when the eigenvalues of $\boldsymbol{\mathcal{F}}$ or $\boldsymbol{\mathcal{U}}$ are all positive. 
It is possible to show that the mass matrix is still diagonalizable with real eigenvalues if there exists a matrix
\begin{equation}
\label{Y_def}
\boldsymbol{\mathcal{Y}} = \sigma \boldsymbol{\mathcal{U}} + \gamma \boldsymbol{\mathcal{F}} \,, \qquad \sigma^2 + \gamma^2 = 1 \,,
\end{equation} 
that is positive definite \cite{citekey}.
For example if $\boldsymbol{\mathcal{F}}$ has all negative eigenvalues then we can choose $\sigma=0$, $\gamma=-1$ and simultaneously diagonalize $-\boldsymbol{\mathcal{F}}$ and $\boldsymbol{\mathcal{U}}$. Alternatively if $\boldsymbol{\mathcal{U}}$ has all negative eigenvalues the good choice would be $\sigma=-1$, $\gamma=0$.
Even if a positive definite matrix $\boldsymbol{\mathcal{Y}}$ does not exist, it may still be possible to diagonalize the mass matrix with complex eigenvalues.

\section{Boundary value problem}
\label{App_B}
In this appendix we discuss the boundary value problem which arises when solving the second order scalar field equation as discussed in section~\ref{subsec:scalo2}. For simplicity, we restrict our discussion here to a single eigenvalue \(\lambda\) of a diagonalizable mass matrix, so that we have to solve an equation of the form
\begin{equation}
\nabla^2\PPh{2} - \lambda\PPh{2} = \hat{k}\rho\,,
\end{equation}
where \(\rho = M_0\delta(r)\) is the matter density of a point mass at the origin. Here we omitted the eigenvector index \((\alpha)\) for brevity. We show that the boundary value problem uniquely determines the solution for the scalar field, and thus also the metric perturbation, unless the mass matrix has a negative eigenvalue, in which case there remains an undetermined constant, which must be fixed by other means.

\subsection{Zero eigenvalue}
\label{App_B_zero_eigenvalue}
In the case of a zero eigenvalue \(\lambda = 0\) of the mass matrix we simply have to solve the Poisson equation
\begin{equation}
\nabla^2\PPh{2} = \hat{k}\rho\,,
\end{equation}
which has the general spherically symmetric vacuum solution (outside the point mass source)
\begin{equation}
\PPh{2} = \frac{c_1}{r} + c_2\,.
\end{equation}
From the boundary condition
\begin{equation}
\lim_{r \to \infty}\PPh{2} = 0
\end{equation}
we immediately obtain \(c_2 = 0\). Note that the solution has a singularity at \(r = 0\), which must be matched with the point mass source. We thus consider a spherical integration volume \(B_R\) of radius \(R\) around the point mass, for which we find
\begin{equation}
\hat{k}M_0 = \hat{k}\iiint_{B_R}\rho\,dV
= \iiint_{B_R}\nabla^2\PPh{2}\,dV
= \oiint_{\partial B_R}\vec{\nabla}\PPh{2} \cdot d\vec{A}
= -c_1\oiint_{\partial B_R}\frac{\vec{e}_r}{r^2} \cdot d\vec{A}
= -4\pi c_1\,.
\end{equation}
Thus, we have
\begin{equation}
c_1 = -\frac{\hat{k}M_0}{4\pi}
\end{equation}
and
\begin{equation}
\PPh{2} = -\frac{\hat{k}M_0}{4\pi r}\,.
\end{equation}
This is of course the classical and well-known solution of the Poisson equation for a pointlike source.

\subsection{Positive eigenvalue}
\label{App_B_pos_eigenvalue}
For a positive eigenvalue \(\lambda = m^2\) with \(m > 0\) we have a screened Poisson equation
\begin{equation}
\nabla^2\PPh{2} - m^2\PPh{2} = \hat{k}\rho\,,
\end{equation}
with the general spherically symmetric vacuum solution
\begin{equation}
\PPh{2} = \frac{c_1e^{-mr} + c_2e^{mr}}{r}\,.
\end{equation}
Also here we obtain \(c_2 = 0\) from the boundary condition that \(\PPh{2}\) vanishes at infinity. We further have
\begin{eqnarray}
\hat{k}M_0 &=& \hat{k}\iiint_{B_R}\rho\,dV
= \iiint_{B_R}(\nabla^2\PPh{2} - m^2\PPh{2})dV
= \oiint_{\partial B_R}\vec{\nabla}\PPh{2} \cdot d\vec{A} - m^2\iiint_{B_R}\PPh{2}\,dV \nonumber \\
&=& -c_1\oiint_{\partial B_R}\frac{(1 + mr)e^{-mr}}{r^2}\vec{e}_r \cdot d\vec{A} - c_1m^2\iiint_{B_R}\frac{e^{-mr}}{r}dV \nonumber \\
&=& -4\pi c_1(1 + mR)e^{-mR} - 4\pi c_1\left[1 - (1 + mR)e^{-mR}\right]
= -4\pi c_1\,.
\end{eqnarray}
We thus have
\begin{equation}
c_1 = -\frac{\hat{k}M_0}{4\pi}
\end{equation}
and
\begin{equation}
\PPh{2} = -\frac{\hat{k}M_0}{4\pi r}e^{-mr}\,.
\end{equation}
The solution is thus given by a Yukawa potential.

\subsection{Negative eigenvalue}
\label{App_B_neg_eigenvalue}
For a negative eigenvalue \(\lambda = -n^2\) with \(n > 0\) we have an inhomogeneous Helmholtz equation
\begin{equation}
\nabla^2\PPh{2} + n^2\PPh{2} = \hat{k}\rho\,,
\end{equation}
with the general spherically symmetric vacuum solution
\begin{equation}
\PPh{2} = \frac{c_1\cos nr + c_2\sin nr}{r}\,.
\end{equation}
In this case we cannot eliminate \(c_2\) using the condition that \(\PPh{2}\) vanishes at infinity. From the point mass source we obtain
\begin{eqnarray}
\hat{k}M_0 &=& \hat{k}\iiint_{B_R}\rho\,dV
= \iiint_{B_R}(\nabla^2\PPh{2} + n^2\PPh{2})dV
= \oiint_{\partial B_R}\vec{\nabla}\PPh{2} \cdot d\vec{A} + n^2\iiint_{B_R}\PPh{2}\,dV \nonumber \\
&=& -\oiint_{\partial B_R}\frac{(c_1 - c_2nr)\cos nr + (c_2 + c_1nr)\sin nr}{r^2}\vec{e}_r \cdot d\vec{A} + n^2\iiint_{B_R}\frac{c_1\cos nr + c_2\sin nr}{r}dV \nonumber \\
&=& -4\pi c_1\,.
\end{eqnarray}
The constant \(c_2\) remains undetermined here. It must be fixed by some additional reasoning, for instance by letting the theory parameters to approach the limit to general relativity and matching the solution to the respective solution in general relativity (cf.\ e.g.\ Ref.~\cite{Kehagias:2014sda}).

\subsection{Complex eigenvalue}
\label{App_B_complex_eigenvalue}
For a complex eigenvalue \(\lambda = (m + in)^2\) we have an equation of the form
\begin{equation}
\nabla^2\PPh{2} - (m + in)^2\PPh{2} = \hat{k}\rho\,,
\end{equation}
with the general spherically symmetric vacuum solution
\begin{equation}
\PPh{2} = \frac{c_1e^{-(m + in)r} + c_2e^{(m + in)r}}{r}
= \frac{c_1e^{-mr}(\cos nr - i\sin nr) + c_2e^{mr}(\cos nr + i\sin nr)}{r}\,.
\end{equation}
Since \(\lambda\) is complex, we can always choose \(m + in\) to be the root of \(\lambda\) which has a real part \(m > 0\). Hence, we obtain \(c_2 = 0\) from the boundary condition that \(\PPh{2}\) vanishes at infinity. We further have
\begin{eqnarray}
\hat{k}M_0 &=& \hat{k}\iiint_{B_R}\rho\,dV
= \iiint_{B_R}\left[\nabla^2\PPh{2} - (m + in)^2\PPh{2}\right]dV
= \oiint_{\partial B_R}\vec{\nabla}\PPh{2} \cdot d\vec{A} - (m + in)^2\iiint_{B_R}\PPh{2}\,dV \nonumber \\
&=& -c_1\oiint_{\partial B_R}\frac{(1 + mr - inr)(\cos nr - i\sin nr)e^{-mr}}{r^2}\vec{e}_r \cdot d\vec{A} \nonumber \\
&\phantom{=}& \qquad - c_1(m + in)^2\iiint_{B_R}\frac{e^{-mr}(\cos nr - i\sin nr)}{r}dV \nonumber \\
&=& -4\pi c_1\,.
\end{eqnarray}
We thus have
\begin{equation}
c_1 = -\frac{\hat{k}M_0}{4\pi}
\end{equation}
and
\begin{equation}
\PPh{2} = -\frac{\hat{k}M_0}{4\pi r}e^{-mr}(\cos nr - i\sin nr)\,.
\end{equation}
The solution is complex. However, recall that in this case there always exists a complex conjugate eigenvalue \(\lambda^* = (m - in)^2\), and that the corresponding eigenvectors are such that its source coupling is given by \(k^*\), so that the field equation is solved by \(\PPh{2}^*\), and that the original field equations in the untransformed basis have the real solutions \(\PPh{2} + \PPh{2}^*\) and \(-i(\PPh{2} - \PPh{2}^*)\).

\section{Non-diagonalizable mass matrix}
\label{App_C}
\label{higer_dim_Jordan_blocks}
Here we extend the treatment of Sec.~\ref{subsec:scalo2} and show how to proceed with solving the scalar field equation in case the mass matrix \(\boldsymbol{\mathcal{M}}\) is non-diagonalizable. Instead of Eq.\ (\ref{Poisson screened nonnegative distinct eigenvalue}), in the general case the scalar equation assumes the form
\begin{equation}\label{eqn:jordanscaleq}
\nabla^2\PPh{2}^{(\alpha)} - J^{(\alpha)}{}_{(\beta)}\PPh{2}^{(\beta)} = \hat{k}^{(\alpha)}\rho\,,
\end{equation}
where \(\mathbf{J}\) is the Jordan normal form of the mass matrix. In the following discussion we will assume that \(\mathbf{J}\) consists of only a single Jordan block. In the most general case, where \(\mathbf{J}\) consists of a direct sum of several Jordan blocks, this discussion should be applied to each block separately. We distinguish three cases, depending on whether the diagonal elements of the Jordan block are positive, zero, or complex. As noted before in Appendix \ref{App_B_neg_eigenvalue}, the case of negative eigenvalue runs into a problem of how to fix the integration constant, and it is a bit complicated to further generalize this case to a nontrivial Jordan block.

\subsection{Positive eigenvalue}
First we consider the case that the Jordan normal form is given by a single block with
\begin{equation}
\mathbf{J} = \mathbf{D} + \mathbf{N}\,, \quad \mathbf{D} = \left(\begin{array}{ccc}
m^2 & & 0\\
& \ddots &\\
0 & & m^2
\end{array}\right)\,, \quad \mathbf{N} = \left(\begin{array}{cccc}
0 & \mathfrak{m}^2 & & 0\\
& \ddots & \ddots &\\
& & \ddots & \mathfrak{m}^2\\
0 & & & 0
\end{array}\right)\,,
\end{equation}
where both \(m\) and \(\mathfrak{m}\) are non-zero and real. Usually one would normalize \(\mathfrak{m}^2 = 1\); however, recall that \(\mathbf{J}\) is of dimension (mass)$^2$, so that we introduce a unit mass here. Note that \(\mathbf{D}\) is diagonal and \(\mathbf{N}\) is nilpotent, \(\mathbf{N}^d = 0\). Further, \(\mathbf{D}\) and \(\mathbf{N}\) commute, \(\mathbf{D}\mathbf{N} = \mathbf{N}\mathbf{D}\), since \(\mathbf{D}\) is a multiple of the unit matrix.

We can now solve the scalar field equations~\eqref{eqn:jordanscaleq} recursively, starting with the last equation \(\alpha = d\), which reads
\begin{equation}
(\nabla^2 - m^2)\PPh{2}^{(d)} = \hat{k}^{(d)}\rho\,.
\end{equation}
We obtain the solution
\begin{equation}
\PPh{2}^{(d)} = -\hat{k}^{(d)}\frac{M_0}{4\pi r}e^{-mr}
\end{equation}
for a point mass \(M_0\) at the origin, making use of the boundary conditions detailed in the preceding Appendix \ref{App_B_neg_eigenvalue}. The remaining field equations then take the form
\begin{equation}
(\nabla^2 - m^2)\PPh{2}^{(\alpha)} = \mathfrak{m}^2\PPh{2}^{(\alpha + 1)} + \hat{k}^{(\alpha)}\rho\,.
\end{equation}
They can be solved by an ansatz of the form
\begin{equation}
\PPh{2}^{(\alpha)} = -\frac{M_0}{4\pi r}e^{-mr}\sum_{\beta = \alpha}^dp_{(\beta - \alpha)}(r)\hat{k}^{(\beta)}\,,
\end{equation}
where \(p_{(\alpha)}\) is determined by the recursive definition
\begin{equation}
p_{(0)}(r) = 1\,, \quad \mathfrak{m}^2p_{(\alpha - 1)}(r) = p_{(\alpha)}''(r) - 2mp_{(\alpha)}'(r)\,.
\end{equation}
The solution which is compatible with the boundary conditions takes the form
\begin{subequations}\label{eqn:pfuncr}
\begin{eqnarray}
p_{(\alpha)}(r) &=& \left(\frac{\mathfrak{m}}{m}\right)^{2\alpha}\sum_{\beta = 0}^{\alpha}\frac{(-1)^{\beta}(\alpha)_{\alpha - \beta}(1 - \alpha)_{\alpha - \beta}}{2^{2\alpha - \beta}\alpha!(\alpha - \beta)!}(mr)^{\beta}\\
&=& \left(\frac{\mathfrak{m}}{m}\right)^{2\alpha}\sum_{\beta = 0}^{\alpha}(-1)^{\alpha}\binom{2\alpha - \beta - 1}{\beta - 1}\frac{(2\alpha - 2\beta)!}{2^{2\alpha - \beta}\alpha!(\alpha - \beta)!}(mr)^{\beta}\\
&=& \begin{cases}
0 & \text{if $\alpha < 0$,}\\
1 & \text{if $\alpha = 0$,}\\
\left(\frac{\mathfrak{m}}{m}\right)^{2\alpha}\frac{(-1)^{\alpha}\Gamma\left(\alpha - \frac{1}{2}\right)}{2\sqrt{\pi}\alpha!} \, \hypgeo{1}{1}(1 - \alpha; 2 - 2\alpha; 2mr)mr & \text{otherwise}\\
\end{cases}\\
&=& \begin{cases}
0 & \text{if $\alpha < 0$,}\\
1 & \text{if $\alpha = 0$,}\\
-\frac{2mr}{4^{\alpha}\alpha}\left(\frac{\mathfrak{m}}{m}\right)^{2\alpha}L_{\alpha - 1}^{(1 - 2\alpha)}(2mr) & \text{otherwise},\\
\end{cases}
\end{eqnarray}
\end{subequations}
where $\Gamma$ denotes the gamma function, $\hypgeo{1}{1}$ represents the confluent hypergeometric function of the first kind, and $L_n^{(a)}$ is the $n^{\textrm{th}}$ generalized Laguerre polynomial.

We could have obtained this solution also from the matrix ansatz
\begin{equation}\label{eqn:ndsolmatr}
\PPh{2}^{(\alpha)} = -\frac{M_0}{4\pi r}E^{(\alpha)}{}_{(\beta)}\hat{k}^{(\beta)}\,,
\end{equation}
where
\begin{equation}
\mathbf{E} = \exp\left(-r\sqrt{\mathbf{J}}\right)\,.\label{eqn:ematr}
\end{equation}
Here we can determine the square root using
\begin{equation}
\sqrt{\mathbf{J}} = \sqrt{\mathbf{D} + \mathbf{N}}
= \sqrt{\mathbf{D}}\sqrt{\mathds{1}_d + \mathbf{D}^{-1}\mathbf{N}}
= \sum_{j = 0}^{d - 1}\frac{\sqrt{\pi}}{2j!\Gamma\left(\frac{3}{2} - j\right)}\mathbf{D}^{\frac{1}{2} - j}\mathbf{N}^j\,,
\end{equation}
while the components are then given by
\begin{equation}
(\sqrt{{J}})^{(\alpha)}{}_{(\beta)} = \frac{\sqrt{\pi}}{2(\beta - \alpha)!\Gamma\left(\frac{3}{2} - \beta + \alpha\right)}\,\frac{\mathfrak{m}^{2(\beta - \alpha)}}{m^{2(\beta - \alpha) - 1}}\,.
\end{equation}
The exponential is defined as usual through the Taylor series
\begin{equation}
\mathbf{E} = \sum_{j = 0}^{\infty}\frac{\left(-r\sqrt{\mathbf{J}}\right)^j}{j!}\,,
\end{equation}
and its components are
\begin{equation}
E^{(\alpha)}{}_{(\beta)} = p_{(\beta - \alpha)}(r) \, e^{-mr}\,,
\end{equation}
where \(p_{(\alpha)}(r)\) is given by the explicit formula~\eqref{eqn:pfuncr}.

\subsection{Zero eigenvalue}
Let us now assume that the Jordan normal form of the mass matrix is given by a single block
\begin{equation}
\mathbf{J} = \mathbf{N} = \left(\begin{array}{cccc}
0 & \mathfrak{m}^2 & & 0\\
& \ddots & \ddots &\\
& & \ddots & \mathfrak{m}^2\\
0 & & & 0
\end{array}\right)\,,
\end{equation}
so that \(\mathbf{D} = 0\) in the notation of the preceding section. Note that \(\mathbf{J}\) is nilpotent, \(\mathbf{J}^d = 0\).

Also in this case we can solve these equations recursively, starting with the last equation \(\alpha = d\), which reads
\begin{equation}
\nabla^2\PPh{2}^{(d)} = \hat{k}^{(d)}\rho\,.
\end{equation}
We obtain the solution
\begin{equation}
\PPh{2}^{(d)} = -\hat{k}^{(d)}\frac{M_0}{4\pi r}
\end{equation}
for a point mass \(M_0\) at the origin, again making use of the boundary conditions detailed in Appendix \ref{App_B_zero_eigenvalue}. The remaining field equations then take the form
\begin{equation}
\nabla^2\PPh{2}^{(\alpha)} = \mathfrak{m}^2\PPh{2}^{(\alpha + 1)} + \hat{k}^{(\alpha)}\rho\,.
\end{equation}
They can be solved by an ansatz of the form
\begin{equation}
\PPh{2}^{(\alpha)} = -\frac{M_0}{4\pi r}\sum_{\beta = \alpha}^dp_{(\beta - \alpha)}(r)\hat{k}^{(\beta)}\,,
\end{equation}
where \(p_{(\alpha)}\) is determined by the recursive definition
\begin{equation}
p_{(0)}(r) = 1\,, \quad \mathfrak{m}^2p_{(\alpha - 1)}(r) = p_{(\alpha)}''(r)\,.
\end{equation}
The solution which is compatible with the boundary conditions takes the form
\begin{equation}
p_{(\alpha)}(r) = \frac{(\mathfrak{m} r)^{2\alpha}}{(2\alpha)!}\,.
\end{equation}

Also in this case we can write the solution as
\begin{equation}
\PPh{2}^{(\alpha)} = -\frac{M_0}{4\pi r}E^{(\alpha)}{}_{(\beta)}\hat{k}^{(\beta)}\,,\label{eqn:ndsolmatz}
\end{equation}
where
\begin{equation}
\mathbf{E} = \sum_{j = 0}^{\infty}\frac{\mathbf{J}^jr^{2j}}{(2j)!}\,.\label{eqn:ematz}
\end{equation}
Note that this sum terminates at \(j = d\), because \(\mathbf{J}\) is nilpotent. Its components are given by
\begin{equation}
E^{(\alpha)}{}_{(\beta)} = \begin{cases}
0 & \text{if $\alpha > \beta$,}\\
\frac{(\mathfrak{m} r)^{2(\beta - \alpha)}}{[2(\beta - \alpha)]!} & \text{otherwise.}
\end{cases}
\end{equation}

\subsection{Complex eigenvalues}
We finally discuss the case that the Jordan normal form is given by a pair of complex conjugate blocks
\begin{equation}
J^{(\alpha)}{}_{(\beta)} = \left(\begin{array}{cccc|cccc}
(m - in)^2 & \mathfrak{m}^2 & & & & & & 0\\
& \ddots & \ddots & & & & &\\
& & \ddots & \mathfrak{m}^2 & & & &\\
& & & (m - in)^2 & & & &\\\hline
& & & & (m + in)^2 & \mathfrak{m}^2 & &\\
& & & & & \ddots & \ddots &\\
& & & & & & \ddots & \mathfrak{m}^2\\
0 & & & & & & & (m + in)^2
\end{array}\right) \,
\end{equation}
where \(m > 0\) and \(n > 0\).
It is convenient to label the components \(1^-, \ldots, d^-, 1^+, \ldots, d^+\). The last field equation in each block then takes the form
\begin{equation}
\left[\nabla^2 - (m \pm in)^2\right]\PPh{2}^{(d^{\pm})} = \hat{k}^{(d^{\pm})}\rho\,.
\end{equation}
Once more making use of the boundary conditions detailed in Appendix \ref{App_B_complex_eigenvalue}, the solution is given by
\begin{equation}
\PPh{2}^{(d^{\pm})} = -\hat{k}^{(d^{\pm})}\frac{M_0}{4\pi r}e^{-(m \pm in)r} = -\hat{k}^{(d^{\pm})}\frac{M_0}{4\pi r}e^{-mr}\left(\cos nr \mp i\sin nr\right)\,.
\end{equation}
The remaining field equations then take the form
\begin{equation}
\left[\nabla^2 - (m \pm in)^2\right]\PPh{2}^{(\alpha^{\pm})} = \mathfrak{m}^2\PPh{2}^{(\alpha + 1)^{\pm}} + \hat{k}^{(\alpha^{\pm})}\rho\,.
\end{equation}
They can be solved by an ansatz of the form
\begin{equation}
\PPh{2}^{(\alpha^{\pm})} = -\frac{M_0}{4\pi r}e^{-(m \pm in)r}\sum_{\beta = \alpha}^dp^{\pm}_{(\beta - \alpha)}(r)\hat{k}^{(\beta^{\pm})}\,,
\end{equation}
where \(p^{\pm}_{(\alpha)}\) is determined by the recursive definition
\begin{equation}
p^{\pm}_{(0)}(r) = 1\,, \quad \mathfrak{m}^2p^{\pm}_{(\alpha - 1)}(r) = {p^{\pm}_{(\alpha)}}''(r) - 2(m \pm in){p^{\pm}_{(\alpha)}}'(r)\,.
\end{equation}
The solution which is compatible with the boundary conditions takes the form
\begin{subequations}\label{eqn:pfuncc}
\begin{eqnarray}
p^{\pm}_{(\alpha)}(r) &=& \left(\frac{\mathfrak{m}}{m \pm in}\right)^{2\alpha}\sum_{\beta = 0}^{\alpha}\frac{(-1)^{\beta}(\alpha)_{\alpha - \beta}(1 - \alpha)_{\alpha - \beta}}{2^{2\alpha - \beta}\alpha!(\alpha - \beta)!}((m \pm in)r)^{\beta}\\
&=& \left(\frac{\mathfrak{m}}{m \pm in}\right)^{2\alpha}\sum_{\beta = 0}^{\alpha}(-1)^{\alpha}\binom{2\alpha - \beta - 1}{\beta - 1}\frac{(2\alpha - 2\beta)!}{2^{2\alpha - \beta}\alpha!(\alpha - \beta)!}((m \pm in)r)^{\beta}\\
&=& \begin{cases}
0 & \text{if $\alpha < 0$,}\\
1 & \text{if $\alpha = 0$,}\\
\left(\frac{\mathfrak{m}}{m \pm in}\right)^{2\alpha}\frac{(-1)^{\alpha}\Gamma\left(\alpha - \frac{1}{2}\right)}{2\sqrt{\pi}\alpha!}\hypgeo{1}{1}(1 - \alpha; 2 - 2\alpha; 2(m \pm in)r)(m \pm in)r & \text{otherwise}\\
\end{cases}\\
&=& \begin{cases}
0 & \text{if $\alpha < 0$,}\\
1 & \text{if $\alpha = 0$,}\\
-\frac{2(m \pm in)r}{4^{\alpha}\alpha}\left(\frac{\mathfrak{m}}{m \pm in}\right)^{2\alpha}L_{\alpha - 1}^{(1 - 2\alpha)}(2(m \pm in)r) & \text{otherwise},\\
\end{cases}
\end{eqnarray}
\end{subequations}
with the functions \(\Gamma\), \(\hypgeo{1}{1}\) and \(L_n^{(a)}\) as above in Eq.~\eqref{eqn:pfuncr}.

Writing this solution in matrix form, we have
\begin{equation}
\PPh{2}^{(\alpha)} = -\frac{M_0}{4\pi r}E^{(\alpha)}{}_{(\beta)}\hat{k}^{(\beta)}\,,\label{eqn:ndsolmatc}
\end{equation}
where again we used the matrix exponential
\begin{equation}
\mathbf{E} = \exp\left(-r\sqrt{\mathbf{J}}\right) = \sum_{j = 0}^{\infty}\frac{\left(-r\sqrt{\mathbf{J}}\right)^j}{j!}\,.\label{eqn:ematc}
\end{equation}
In this case the components are
\begin{equation}
(\sqrt{{J}})^{(\alpha^{\pm})}{}_{(\beta^{\pm})} = \frac{\sqrt{\pi}}{2(\beta - \alpha)!\Gamma\left(\frac{3}{2} - \beta + \alpha\right)}\frac{\mathfrak{m}^{2(\beta - \alpha)}}{(m \pm in)^{2(\beta - \alpha) - 1}}\,, \quad (\sqrt{{J}})^{(\alpha^{\pm})}{}_{(\beta^{\mp})} = 0
\end{equation}
and thus
\begin{subequations}
\begin{eqnarray}
E^{(\alpha^{\pm})}{}_{(\beta^{\pm})}(r) &=& p^{\pm}_{(\beta - \alpha)} \, e^{-(m \pm in)r}\,,\\
E^{(\alpha^{\pm})}{}_{(\beta^{\mp})} &=& 0\,,
\end{eqnarray}
\end{subequations}
with \(p^{\pm}_{(\alpha)}(r)\) given by the defining formula~\eqref{eqn:pfuncc}.

\subsection{General solution formula}
We have seen in the previous sections that the solutions~\eqref{eqn:ndsolmatr},~\eqref{eqn:ndsolmatz},~\eqref{eqn:ndsolmatc} for the second order scalar field equations can always be written in a common matrix form
\begin{equation}
\PPh{2}^{(\alpha)} = -\frac{M_0}{4\pi r}E^{(\alpha)}{}_{(\beta)}\hat{k}^{(\beta)}\,,
\end{equation}
with a matrix \(\mathbf{E}\) given by equations~\eqref{eqn:ematr},~\eqref{eqn:ematz},~\eqref{eqn:ematc}. Note that these formulas are very similar, and can be brought to a common form
\begin{equation}
\mathbf{E} = \sum_{j = 0}^{\infty}\left(\frac{\mathbf{J}^jr^{2j}}{(2j)!} - \frac{\mathbf{S}^{2j + 1}r^{2j + 1}}{(2j + 1)!}\right)\,, \quad \mathbf{S} = \begin{cases}
\sqrt{\mathbf{J}} & \text{if $\mathbf{J}$ has a square root,}\\
0 & \text{otherwise.}
\end{cases}
\end{equation}
Here \(\sqrt{\mathbf{J}}\) always denotes the positive square root, i.e., the square root with positive real parts of the eigenvalues. Note that this formula holds only for a single Jordan block. In case that \(\mathbf{J}\) consists of several Jordan blocks \(\tilde{\mathbf{J}}\), and is thus given by a direct sum
\begin{equation}
\mathbf{J} = \bigoplus_{\text{Jordan blocks $\tilde{\mathbf{J}}$}}\tilde{\mathbf{J}}\,,
\end{equation}
it must be applied to each Jordan block separately. The full matrix \(\mathbf{E}\) is then given likewise by a direct sum
\begin{equation}
\label{final_formula}
\mathbf{E} = \bigoplus_{\text{Jordan blocks $\tilde{\mathbf{J}}$}}\sum_{j = 0}^{\infty}\left(\frac{\tilde{\mathbf{J}}^jr^{2j}}{(2j)!} - \frac{\tilde{\mathbf{S}}^{2j + 1}r^{2j + 1}}{(2j + 1)!}\right)\,, \quad \tilde{\mathbf{S}} = \begin{cases}
\sqrt{\tilde{\mathbf{J}}} & \text{if $\tilde{\mathbf{J}}$ has a square root,}\\
0 & \text{otherwise.}
\end{cases}
\end{equation}


\end{document}